\documentclass[conference]{IEEEtran}

\usepackage{amsmath}
\usepackage{xcolor}
\usepackage{soul}
\usepackage[braket, qm]{qcircuit}
\usepackage{physics}
\usepackage{tikz}
\def\checkmark{\tikz\fill[scale=0.4](0,.35) -- (.25,0) -- (1,.7) -- (.25,.15) -- cycle;} 
\usepackage{filecontents,lipsum}
\usepackage[noadjust]{cite}
\usepackage{enumitem}
\usepackage{multirow}
\usepackage{booktabs}
\usepackage{algorithmicx}
\usepackage{algpseudocode}
\usepackage{algorithm}

\usepackage{caption}
\usepackage{subcaption}

\usepackage{balance}
\usepackage{nicematrix}
\usepackage{hyperref}

\DeclareMathOperator*{\argmin}{argmin}

\begin{document}

\title{Exploration of Quantum Computer Power Side-Channels}

\author{
\IEEEauthorblockN{Chuanqi Xu}
\IEEEauthorblockA{\textit{Electrical Engineering} \\
\textit{Yale University}\\
New Haven, CT, USA \\
chuanqi.xu@yale.edu}
\and
\IEEEauthorblockN{Ferhat Erata}
\IEEEauthorblockA{\textit{Computer Science} \\
\textit{Yale University}\\
New Haven, CT, USA \\
ferhat.erata@yale.edu}
\and
\IEEEauthorblockN{Jakub Szefer}
\IEEEauthorblockA{\textit{Electrical Engineering} \\
\textit{Yale University}\\
New Haven, CT, USA \\
jakub.szefer@yale.edu}
}

\maketitle
\begin{abstract}
    With the rapidly growing interest in quantum computing also grows the importance of securing these quantum computers from various physical attacks. Constantly increasing qubit counts and improvements to the fidelity of the quantum computers hold great promise for the ability of these computers to run novel algorithms with highly sensitive intellectual property. However, in today's cloud-based quantum computer setting, users lack physical control over the computers. Physical attacks, such as those perpetrated by malicious insiders in data centers, could be used to extract sensitive information about the circuits being executed on these computers. Indeed, this work shows for the first time that power-based side-channel attacks could be deployed against quantum computers. Such attacks can be used to recover information about the control pulses sent to these computers. By analyzing these control pulses, attackers can reverse-engineer the equivalent gate-level description of the circuits, and possibly even the secret algorithms being run. This work introduces five new types of attacks, and evaluates them using control pulse information available from cloud-based quantum computers. This work demonstrates how and what circuits could be recovered, and then in turn how to defend from the newly demonstrated side-channel attacks on quantum~computers.
\end{abstract}

\section{Introduction}

Quantum computers have gained more and more attention in recent years, especially as large numbers of quantum computers are now easily accessible over the internet. 
Cloud-based vendors such as IBM Quantum~\cite{ibm_quantum}, Amazon Bracket~\cite{braket}, and Microsoft Azure~\cite{azure}, already provide access to various types of Noisy Intermediate-Scale Quantum (NISQ) {devices} from different providers. Remote access makes it easy for different users and companies to run algorithms on real quantum computers without the need to purchase or maintain them. Already, a large number of companies and startups are working on the development of quantum algorithms to run on these cloud-based quantum computers. These companies or startups do not themselves have quantum computers, but depend on remote access to real machines from the cloud providers. They can use a convenient pay-per-use model to run circuits on real quantum computers. However, given possibly important intellectual property embedded in their quantum circuits, there is a need to understand if and how sensitive information could be extracted from the operational behavior of quantum computers. 

Especially, these users, startups, or companies have no control over the physical space where the quantum computers are. While the cloud providers may not be bad actors themselves, the threat of malicious insiders within data centers or cloud computing facilities is well-known in classical security. 
In classical computers, side-channels of different types are a well-known threat~\cite{szefer2018principles}. Among the side-channels, there are timing- and power-based channels, which are major categories of side-channels that have been researched. There are also thermal, EM, acoustic, and a variety of other categories of side-channels.
Timing side-channels are easier to exploit as they only require timing measurement of the victim to be done. Power side-channels are more powerful, but require physical access. With physical access, malicious insiders or other attackers can get detailed information about the execution of the target computer. 

In quantum computers, directly copying the quantum states is not possible due to the no-cloning theorem. The no-cloning theorem states that it is impossible to create an independent and identical copy of an arbitrary unknown quantum state~\cite{park1970concept, wootters1982single, dieks1982communication}. However, there is no such limitation on the classical control operations performed on quantum computers. Quantum computers, such as superconducting qubit machines from IBM, Rigetti, or others, use RF pulses to ``execute'' gate operations on single qubits or two-qubit pairs. The control pulses are fully classical and could be spied on. Given control pulse information, as this work shows, it is possible to reverse engineer the sequence of quantum gates executed on the quantum computer. From the sequence of gates, the algorithm executed can possibly be recovered. As this work shows for the first time, anybody with access to power measurements of the control pulse generation logic can capture and recover the control information. While this work explores power-based side-channels, the same or similar ideas could apply to EM or other types of physical side-channels.

\begin{figure*}[ht]
 \centering
 \hspace{-0.25cm}
 \centering
 \begin{subfigure}[t]{0.48\textwidth}
   \includegraphics[width=\columnwidth]{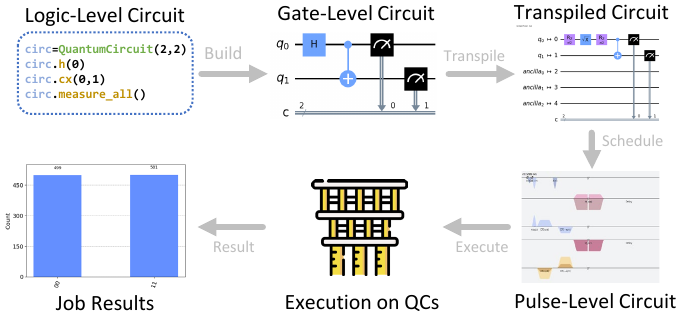}
   \caption{\small Running quantum programs with Qiskit on IBM Quantum.}
   \label{fig:qiskit_process}
 \end{subfigure}
 ~\hspace{0.5cm}
 \begin{subfigure}[t]{0.48\textwidth}
   \includegraphics[width=\columnwidth]{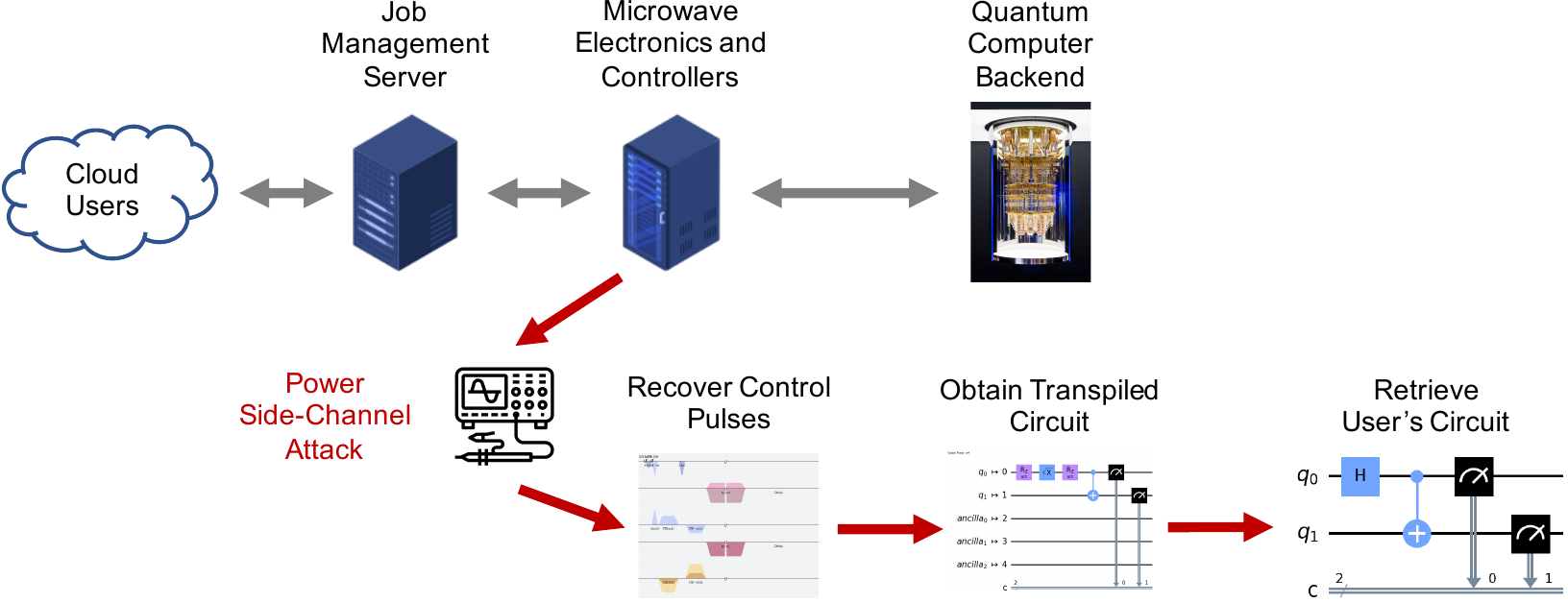}
   \caption{\small Operation of a cloud-based quantum computer, highlighting potential side-channel threats.}
   \label{fig_side_channel_threats}
 \end{subfigure}
 \caption{Process of running quantum circuits using Qiskit on IBM Quantum and the proposed threat model in the process.}
\end{figure*}

In this work, we focus on and demonstrate potential new, side-channels used to extract information about user circuits, i.e., quantum programs. Rather than target the superconducting qubits themselves (which are isolated in a cryogenic refrigerator), we focus on the controller electronics shown in the middle of Figure~\ref{fig_side_channel_threats}.
We note that in the threat model, discussed in more detail in Section~\ref{threat_model}, we assume that the classical computer components, e.g., the job management server, are protected from side-channels. There is a large body of research on the protection of classical computers from power side-channels, e.g., ~\cite{prouff2013masking, synthesis2021icse, synthesis2021date, synthesis2020tornado, synthesis2019tcad, synthesis2019fse, synthesis2018taco, synthesis2017post, synthesis2015tc, synthesis2014cav, synthesis2012ches}. Meanwhile, controller electronics of quantum computers have not been analyzed for potential side-channels before this work.

\subsection{Potential Attacks on Quantum Circuits}
\label{sec:attacker_goals}

The focus of this work is to demonstrate that it can be possible to recover various information about user circuits, i.e., quantum programs, from side-channel information. We present different types of possible information that can be recovered, these can be also considered goals for the attacker:

\begin{enumerate}[label=,itemindent=-2em]
  \item \textbf{(UC) User Circuit Identification} -- Given knowledge about the set of possible circuits executed on the quantum computer, find which circuits the user actually executed.
  \item \textbf{(CO) Circuit Oracle Identification} -- Given a known circuit, such as Bernstein-Vazirani~\cite{doi:10.1137/S0097539796300921}, but an unknown oracle, find the configuration of the oracle used in that circuit.
  \item \textbf{(CA) Circuit Ansatz Identification} -- Given a known circuit, such as a variational circuit used in machine learning applications~\cite{peruzzo2014variational}, but an unknown ansatz, find the configuration of the ansatz used in that circuit.
  \item \textbf{(QM) Qubit Mapping Identification} -- Given a known circuit, identify the placement of which physical qubits were used.
  \item \textbf{(QP) Quantum Processor Identification} -- Given knowledge about the pulses for quantum processors and a circuit, find the quantum processor on which the circuit was executed.
  \item \textbf{(CR) Circuit Reconstruction} -- Given knowledge about the pulses for quantum computer basis gates, reconstruct the complete, unknown circuit from the power traces.
\end{enumerate}

\noindent Considering the attacker's physical access to the quantum computers, this work demonstrates various types of attacks that can be used to recover the above information:

\begin{enumerate}[label=,itemindent=-2em]

  \item \textbf{Timing Attack} -- While this work mainly focuses on power side-channels, we start off by demonstrating simple timing side-channels to help recover user circuits (UC). The limitation of this attack also motivates work on the other power side-channels attacks.

  \item \textbf{Total Energy Attack} -- We next demonstrate that measurement of total energy data can be used to recover users' circuits (UC) as well. This can also be applied to other attackers' goals we listed earlier.

  \item \textbf{Mean Power Attack} -- We also demonstrate a different single measurement attack by showing that measurement of mean power can also be used to recover users' circuits (UC) as well. This can also be applied to other attackers' goals we listed earlier. 

  \item \textbf{Total Power Single Trace Attack} -- A more powerful attacker can measure traces of the total power of all the channels, such attackers can recover user circuits (UC), circuit oracle (CO), circuit ansatz (CA), qubit mapping (QM), and quantum processor (QP) with some accuracy.

  \item \textbf{Per-Channel Power Single Trace Attack} -- Most powerful attackers can collect power traces from channels separately. There are unique drive and control channels, to which microwave pulses are sent, for each single qubit gate and multi-qubit gate. We show that attackers who can collect power traces of these channels can perform circuit reconstruction (CR), thus recovering user circuits.

\end{enumerate}

\section{Background}
\label{sec::background}

This section provides background on quantum computers and typical quantum computer workflow.

\subsection{Qubits and Quantum States}

The quantum bit, or qubit for short, is the most fundamental building block of quantum computing and is conceptually similar to the bit in present classical computing. A qubit, analogous to a bit, has two basis states, denoted by the bra-ket notation as $\ket 0$ and $\ket 1$. However, a qubit can be any linear combination of $\ket 0$ and $\ket 1$ with norm 1, but a classical bit can only be either 0 or 1. Generally, a qubit $\ket \psi$ is more specifically represented as:
\begin{equation*}
  \ket \psi = \alpha \ket 0 + \beta \ket 1,
\end{equation*}
where $\alpha$ and $\beta$ are complex numbers satisfying $|\alpha|^2 + |\beta|^2 = 1$.

It is common to denote qubits using vector representation. The basis states for one qubit can be expressed as two-dimensional vectors, for example, $\ket 0 = [1, 0]^T$ and $\ket 1 = [0, 1]^T$. As a result, the state $\ket \psi$ above can be written as $\ket \psi = \alpha \ket 0 + \beta \ket 1 = [\alpha, \beta]^T$ For multi-qubit states, similar representations exist. For instance, the four basis states $\ket{00}$, $\ket{01}$, $\ket{10}$, and $\ket{11}$ make up the space on which two-qubit states live. More generally, there are $2^n$ basis states in the space of $n$-qubit states, ranging from $\ket{0\dots 0}$ to $\ket{1\dots 1}$, and a $n$-qubit state $\ket \phi$ can be expressed by:
\begin{equation*}
  \ket \phi = \sum_{i = 0}^{2^n - 1} a_i \ket i
\end{equation*}
where $\sum_{i = 0}^{2^n - 1}|a_i|^2 = 1$.

\subsection{Quantum Gates}

Analogous to classical computing, the basic quantum operations at the logic-level are quantum gates. Quantum gates are unitary operations that modify the input qubits, and quantum algorithms consist of a series of quantum gates that can change input qubits into specific quantum states. A quantum gate $U$ must satisfy the equation $U U^\dagger = U^\dagger U = I$, meaning that a quantum gate must be a unitary operation. A quantum gate $U$ operating on a qubit $\ket \psi$ can be written down as $\ket \psi \rightarrow U \ket \psi$. In the vector-matrix representation, $2^n \times 2^n$ matrices can be used to express $n$-qubit quantum gates. For instance, the Pauli-$X$ gate, a single-qubit gate that flips $\ket 0$ to $\ket 1$ and $\ket 1$ to $\ket 0$, is comparable to the NOT gate in classical computation. One another important example is the CNOT gate, also known as the {\tt CX} gate, which is a two-qubit gate that if the control qubit is in the state $\ket 1$, a Pauli-$X$ gate will be applied to the target qubit, and otherwise nothing will happen. Their matrix representations together with some other matrices of quantum gates are shown below. One thing to note is that we follow Qiskit's~\cite{Qiskit} qubit order, where the leftmost qubit is the most significant and the rightmost qubit is the least significant. In light of this, the {\tt CX} gate may have a different matrix representation in other papers if different qubit order is followed:
\begin{equation*}
  {\tt I}=\begin{bmatrix}
    1 & 0 \\
    0 & 1
  \end{bmatrix},\
  {\tt X}=\begin{bmatrix}
    0 & 1 \\
    1 & 0
  \end{bmatrix},\
  {\tt CX} = \begin{bmatrix}
    1 & 0 & 0 & 0 \\
    0 & 0 & 0 & 1 \\
    0 & 0 & 1 & 0 \\
    0 & 1 & 0 & 0
  \end{bmatrix}
\end{equation*}
\begin{equation*}
  {\tt RZ(\theta)}=\begin{bmatrix}
    e^{-i\frac{\theta}{2}} & 0                     \\
    0                      & e^{i\frac{\theta}{2}}
  \end{bmatrix},\
  {\tt SX}=\frac{1}{2}\begin{bmatrix}
    1+i & 1-i \\
    1-i & 1+i
  \end{bmatrix}
\end{equation*}

It has been demonstrated that any unitary quantum gate can be approximated within a minor error using only a small number of quantum gates~\cite{doi:10.1098/rspa.1995.0065}. Therefore, currently available quantum computers usually have a few basis gates, and by grouping the basis gates, they can form other quantum gates. It is not necessary and not possible for them to support all quantum gates. These basis gates, also called native gates, are one of the important configurations of quantum processors. Depending on the low-level control, different manufacturers or even different versions of quantum processors may have different native gates, which is a trade-off between many properties such as error rate and efficiency. In this paper, we based our experiments on IBM Quantum. As an example, for the majority of IBM Quantum quantum computers, the basis gates include {\tt I}, {\tt RZ}, {\tt SX}, {\tt X}, and {\tt CX}. The matrix representations of these gates were shown above this paragraph. Before being run on the actual quantum computing hardware, other quantum gates, like the widely used Hadamard gate, must be decomposed into these basis gates.

\subsection{Control Pulses}
\label{sec::superconducting_quantum_computer_controls}

\begin{figure}[t]
  \centering
  \begin{subfigure}[t]{0.20\textwidth}
    \includegraphics[width=\columnwidth]{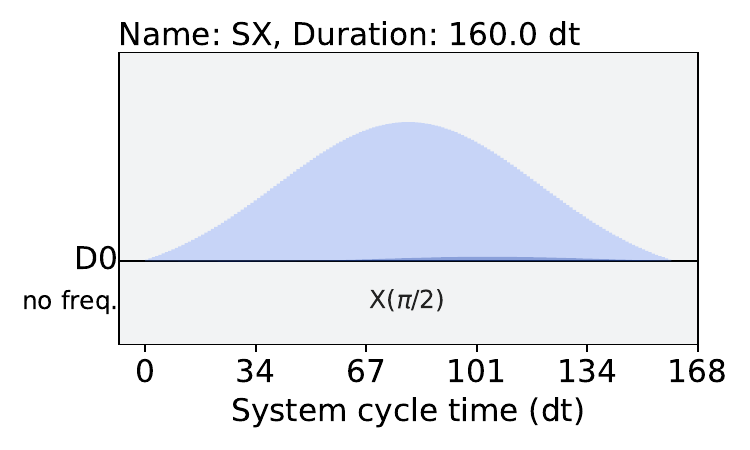}
      \caption{{\tt SX} pulse}
  \end{subfigure}
  ~
  \begin{subfigure}[t]{0.20\textwidth}
    \includegraphics[width=\columnwidth]{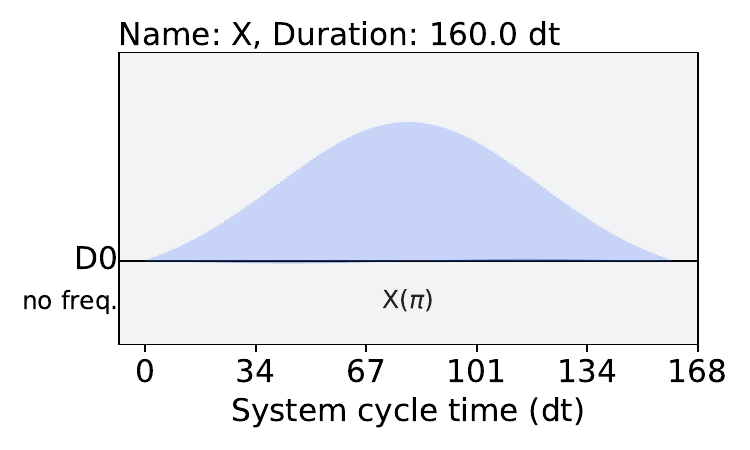}
      \caption{{\tt X} pulse}
  \end{subfigure}
  \begin{subfigure}[t]{0.40\textwidth}
    \includegraphics[width=\columnwidth]{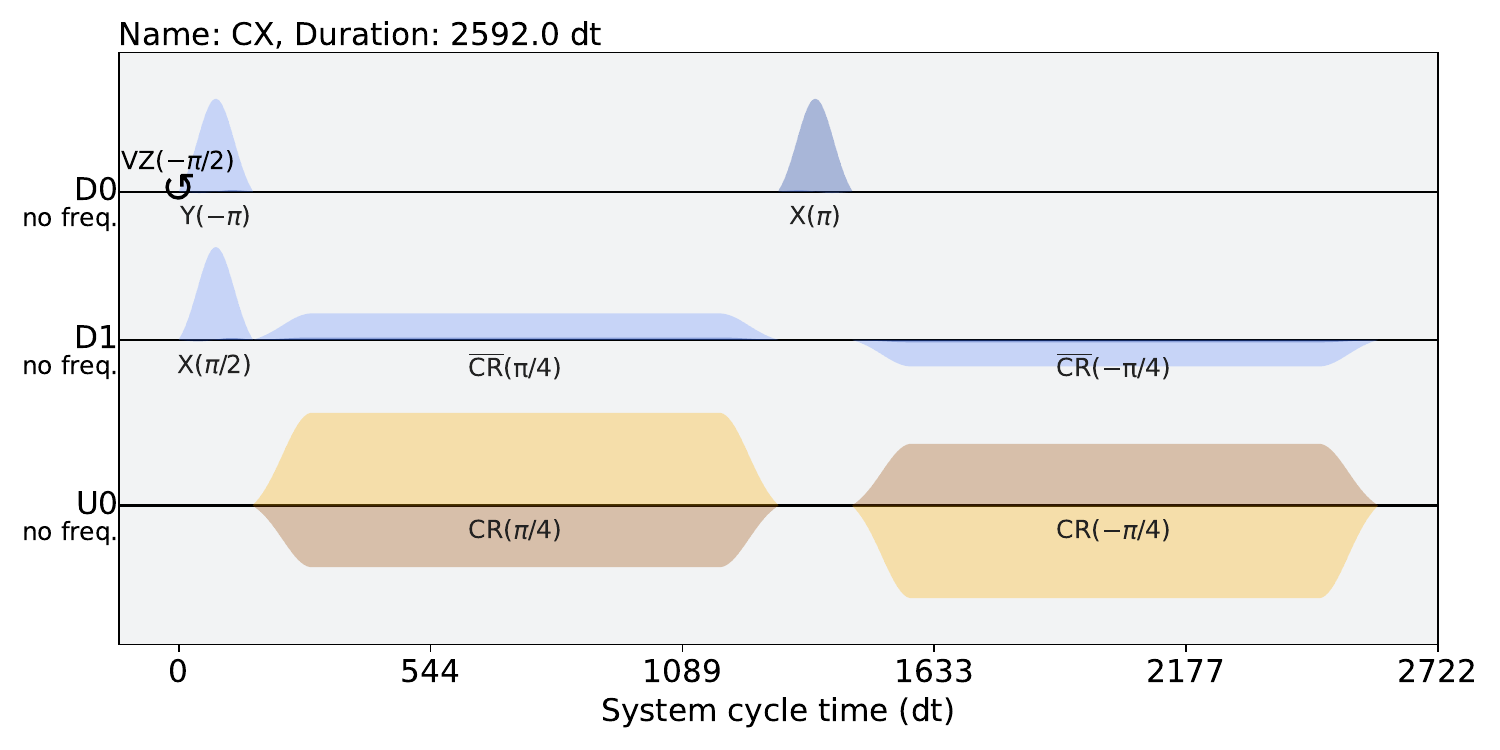}
      \caption{{\tt CX} pulse}
  \end{subfigure}
  \caption{\small {\tt SX}, {\tt X}, and {\tt CX} control pulses. All of the pulses are gathered on {\tt ibm\_lagos}. {\tt SX} and {\tt X} are on qubit 0, and {\tt CX} is on qubit 0 and 1.}
  \label{fig_basis_gates_pulses_example}
\end{figure}

Superconducting qubits are usually controlled by microwave pulses. To actually perform each basis gate on a quantum computer, correct control pulses corresponding to each of the gates need to be generated and sent to the quantum computer. Examples of control pulses for {\tt SX}, {\tt X}, and {\tt CX} gates are shown in Figure~\ref{fig_basis_gates_pulses_example}. On IBM Quantum, {\tt I} gate does nothing and it only adds delays in the control pulses. {\tt RZ} gate is a virtual gate and does not have any real pulse. More details about the virtual {\tt RZ} gate will be discussed in Section~\ref{sec:defense_rz}.

A pulse is usually defined by the envelope, frequency, and phase. As an instance for the superconducting qubit control, the envelope specifies the shape of the signal which is generated by the arbitrary waveform generator (AWG), a common lab instrument, and the frequency and phase specify a period signal that will be used to modulate the envelope signal. These two signals together form the output signal that will be sent to the qubit. The typical settings to drive the qubits are shown in Figure~\ref{fig_hardware_sketch}.

To store envelopes, they are usually discretized into a series of time steps and each element specifies the amplitude at a specific time step. Though envelopes can be in any arbitrary pattern, they are usually parametrized by some predefined shapes so that only a few parameters are needed to specify the envelope. These parameters typically include the duration indicating the length of the pulse, the amplitude indicating the relative strength of the pulse, and other parameters specifying the shape of the pulse. For example, the Derivative Removal by Adiabatic Gate (DRAG) pulse~\cite{PhysRevLett.103.110501, PhysRevA.83.012308} is a standard Gaussian pulse with an additional Gaussian derivative component and lifting applied, and it can be specified with sigma that defines how wide or narrow the Gaussian peak is, and beta that defines the correction amplitude, as well as the duration and amplitude. Another example is the Gaussian square pulse which is a square pulse with a Gaussian-shaped risefall on both sides lifted such that its first sample is zero. Apart from the duration and amplitude, it is parametrized by sigma which defines how wide or narrow the Gaussian risefall is, the width that defines the duration of the embedded square pulse, and the ratio of each risefall duration to~sigma.

On IBM Quantum, the pulses for all native gates are predefined while their parameters are frequently updated by calibrations so that they can maintain high fidelity over time. Pulse parameters are automatically measured and calibrated, and are ready to be used to generate the control pulses for quantum circuits.

\subsection{Pulse-Level Circuit Description}

To fully describe a quantum program, all pulses that need to be performed, when pulses should start relative to the starting point of the circuits, to what qubits the pulses will be applied, and other physical operations like frequency or phase change, need to be specified. This information together with other useful information forms a so-called {\em pulse-level circuit}.

Similar to how pulses are discretized, circuits are also discretized in time steps at the low-level. In this way, pulses can be conveniently fit into the circuits. In addition, it is also necessary to specify to which qubits quantum gates, measurements, and other operations should be applied. With all this information at hand, circuits can be well-defined and ready to be executed in quantum devices. After quantum circuits start to run, when the specified starting time steps are reached, the superconducting quantum computer control equipment sends the pulses defined by their information along electric lines to control the specified qubits.

\subsection{Running Quantum Programs on Quantum Computers}
\label{sec:running_quantum}

To start the process of running a quantum program on nowadays cloud-based superconducting quantum computers, the quantum circuits that solve the desired problem need to be created first. Then the quantum circuits go through a series of transforming processes, and are sent to the cloud to execute and finally users can get the results. We show a typical process of running quantum programs with Qiskit on IBM Quantum in Figure~\ref{fig:qiskit_process}.

The first step is to build the logic-level circuit with a quantum development kit, such as Qiskit~\cite{Qiskit}, Braket SDK~\cite{braket_sdk}, Q\#~\cite{microsoft_q},~Cirq~\cite{cirq_developers_2022}.

The logic-level circuit can also be represented graphically, as shown in the ``Gate-Level Circuit" in Figure~\ref{fig:qiskit_process}, lines that go from left to right stand in for qubits, while the symbols on the lines stand for operations. Without further information, qubits are typically thought to be in the $\ket 0$ state at the start of the quantum circuit. Qubits then evolve through left-to-right sequential processes and are controlled by quantum or classical operations denoted in the circuit plot. For the most part, measurements are performed at the end of the quantum circuit to measure, obtain, and store qubit data in classical memory for future evaluations.

Analogous to classical computing, quantum circuits are usually high-level instructions. Before executing the quantum circuits on quantum computers in reality, a series of operations need to be done to transform them into low-level and hardware-specific instructions, which is similar to the preprocessing, compilation, and assembly process for classical computing programs. To be specific, quantum circuits can be described using a number of different input methods and gates, but eventually, need to be converted to only the native gates supported by the quantum computer.

{\em Transpile} is the term used by Qiskit to stand for the operations and transformations that are like preprocessing and compilation. The process of transpiling involves many steps, including decomposing non-native quantum gates into groups of native gates, grouping and removing quantum gates to reduce the number of gates, mapping the logic qubits in the original circuits to the physical qubits on the specified quantum computers, routing the circuit under limited topologies, potentially optimizing circuits to lower error, and so on. After transpilation, circuits are modified based on hardware-specific knowledge and will generate the same logical results as the original circuits. Circuits up to this point are all gate-level circuits, which use a more general description so that they are understandable by people and can be portable in many cases, though they may still need to be transpiled if they are going to be performed on other kinds of quantum computers.

A lower-level step after transpilation is termed {\em schedule} in Qiskit. Scheduling further maps quantum circuits to microwave pulses, which are the ultimate physical operations used to regulate and control qubits. Because of this, scheduling transforms gate-level circuits into pulse-level circuits. Each microwave pulse is characterized by a series of parameters, such as amplitude and frequency, etc., discussed previously in Section~\ref{sec::superconducting_quantum_computer_controls}. Based on previously calibrated data for each basis gate on each qubit or qubit pair and quantum gadget, scheduling creates microwave pulse sequences. Wave envelopes, frequencies, amplitudes, durations, and other parameters that characterize microwave pulses are included in the data. The final data contains all information that needs to be known by quantum computers to execute the program. After the quantum program starts, the equipment of quantum computers will be manipulated by this information, and qubits are controlled by the equipment to carry out quantum programs.

The steps discussed above convert the initial quantum circuits to a set of instructions that can be used to accomplish the specified quantum programs. As an example of running quantum programs, IBM Quantum provides Qiskit for users as the tool to design circuits, perform these steps, and submit quantum programs to the cloud, and finally, the cloud will execute the users' programs and return the results to users. The above-discussed process needs to be observed in general. In this process, scheduling and even transpilation can be omitted on the user side to simplify the overall development cycles, but they still need to be done on the server side.

\begin{figure*}[th!]
  \centering
  \includegraphics[width=0.8\linewidth]{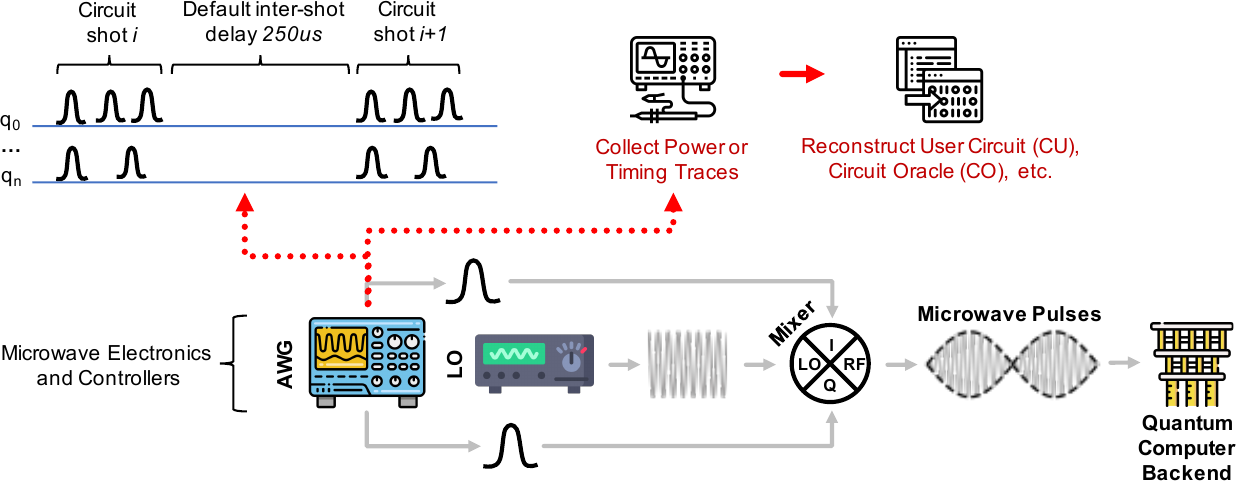}
  \caption{\small Schematic with details of typical qubit drive setup. The local oscillator (LO) generates a low phase-noise microwave carrier signal, and then the wave is modulated in the IQ mixer by I and Q components generated by the arbitrary wave generator (AWG). The pulse is then sent to drive the qubits in the quantum computer. The red line shows the process to collect power traces and timing traces by the attacker. The power traces also can reveal timing information by observing when the control pulses are occurring, as shown in the figure.}
  \label{fig_hardware_sketch}
\end{figure*}

\subsection{Execution of Circuits and Shots}
\label{sec:shot}

In nowadays quantum computing cloud platforms, quantum programs are usually submitted and executed in a particular pattern according to the platform settings. Because the results of most of the quantum algorithms are probabilistic, the same quantum algorithms usually need to be run many times to get the probabilistic results. One execution of the circuit is also often called one shot.

On IBM Quantum, users can submit one circuit or a list of circuits, and specifies how many shots the quantum jobs should run. When the quantum job starts, the quantum circuits are executed one by one sequentially. Between the quantum circuits, there is some reset mechanism to let qubits return to $\ket 0$ states, such as a long duration for qubits to decohere.

\section{Treat Model}
\label{threat_model}

\begin{figure}[t]
     \centering
         \includegraphics[width=0.8\linewidth,trim={0cm 0cm 0cm 0cm},clip]{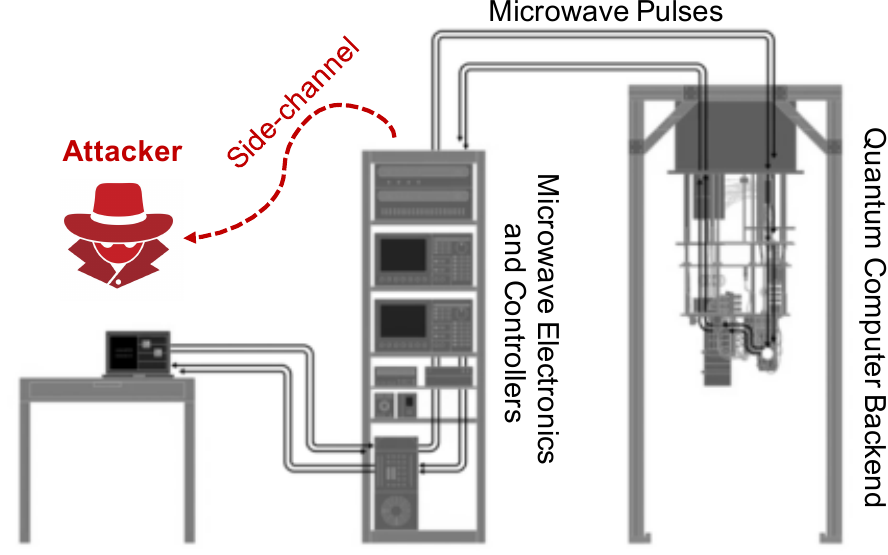}
        \caption{\small Schematic of a typical superconducting quantum computer showing an attacker collecting side-channel information.}
        \label{fig_threat_model}
\end{figure}

The side-channel threat model is depicted in Figure~\ref{fig_threat_model}. More details are shown in Figure~\ref{fig_hardware_sketch}, where the typical qubit drive setup is also illustrated in the figure.

\subsection{Threat Model Background}
\label{sec_definitions}

\subsubsection{Channel}

As introduced in Section~\ref{sec::superconducting_quantum_computer_controls}, pulses are applied to drive designated qubits. Which qubits should be controlled are specified by {\em channels}. Normally there is one channel for single-qubit gates and several channels for multiple-qubit gates. Channels can be mainly categorized into 4 types: drive channels that transmit signals to qubits that enact gate operations, control channels that provide supplementary control over the qubit to the drive channel, measure channels that transmit measurement stimulus pulses for readout, and acquire channels that are used to collect data. Drive channels and control channels are of more interest in this paper because they specify quantum gates. Generally speaking, drive channels correspond to qubits, and control channels correspond to connections between qubits. The number of channels of a quantum device is determined by its architecture. More specifically, the number of drive channels is usually equal to the number of qubits, and the number of control channels is usually equal to the number of connections between two qubits.

\subsubsection{Basis Pulse}

Every quantum circuit needs to be transpiled to a quantum circuit that contains only the basis gates of the target quantum device. We refer to the set of pulses after a basis gate is scheduled as its {\em basis pulses}. Because pulse parameters are highly dependent on qubit physical properties, while the quantum gate is an abstract concept, the same type of gate on different channels has different pulse parameters. For example, {\tt X} gate on qubit $0$ has different pulse parameters from {\tt X} gate on qubits other than $0$. 

\subsubsection{Basis Pulse Library}
\label{sec::basis_pulse_library}

 The set of basis pulses of all basis gates is needed for scheduling. We refer to the set of pulses that defines all basis gates as {\em basis pulse library}. The information on basis pulses is provided by IBM Quantum for all their quantum devices. Notice that IBM Quantum also supports the so-called {\em custom pulse gates}, which allows users to perform gates calibrated with arbitrary pulses~\cite{ibm_custom_gate}, and these gates are not changed in the transpilation and scheduling process. However, 
 for most use cases, custom pulse gates are not needed. Therefore, in our work, we assume that the victim circuits do not contain any custom pulse gates.

\subsubsection{Power Trace}
\label{sec:threat_model_power_trace}

Because pulses are needed to control superconducting qubits, these operations consume energy. We denote {\em power trace} as the time series of the power consumed by the operations controlling qubits. The {\em total power trace} means the time series of the summation of the powers over all channels in a time period, while the {\em per-channel power trace} means the power trace on one specific channel. 
The power consumption of controlling of quantum gates is related to their RF pulses. More generally, we refer to {\em in-channel} and {\em across-channel} as the functions for computing the per-channel power traces and the total power traces from pulse information, respectively. The in-channel function, which we denote as $Power_c[p_c(x)]$, where $c$ represents the channel and $p_c(x)$ represents the pulse amplitude time series on that channel, specifies how the per-channel power traces are computed from pulse amplitudes. The across-channel function, which we denote as $Total[f_{c_1}(x), \dots, f_{c_n}(x)]$, where $c_i, i \in \{1, \dots, n\}$ represent all the channels of one quantum processor, specifies how the total power traces are summed up from all per-channel power traces $f_c(x)$. Based on these definitions, the total power traces $P(x)$ can be computed from the per-channel pulse amplitude time series $p_{c_i}(x)$:
 \begin{equation}
     P(x) = Total\left\{ Power_{c_1}[p_{c_1}(x)], \dots, Power_{c_n}[p_{c_n}(x)] \right\}
 \end{equation}

\subsection{Assumptions of Attacker Measurement}

We assume the attacker can measure timing, power, or energy properties for each shot of a circuit, or they can measure a number of shots and it is easy to divide this into individual shots as discussed below, since all shots perform the same operations. Recall in Section~\ref{sec:shot}, that each quantum program, i.e., quantum circuit, is executed multiple times, and each execution is called a {\em shot}. 

\subsubsection{Per-Shot Timing Measurements}

For the weakest attacker, we assume the attacker is able to measure the execution timing of the victim circuit. As shown in Figure~\ref{fig_hardware_sketch}, we assume the attacker is able to capture the traces of the control pulses. From the traces, the attacker can observe when pulses are occurring. In particular, the shots of a circuit are separated by inter-shot delay, which is used to reset the state of the qubits to $\ket 0$ before the next shot of a circuit is executed. Today this delay in superconducting qubit machines is on the order of 250 us, but will become longer as the decoherence times of the machines increase. The clear separation and the same pattern of the shots allow the attacker to measure their duration, and when one shot ends and the next begins.

\begin{figure*}[t]
  \centering
  \includegraphics[width=0.75\linewidth]{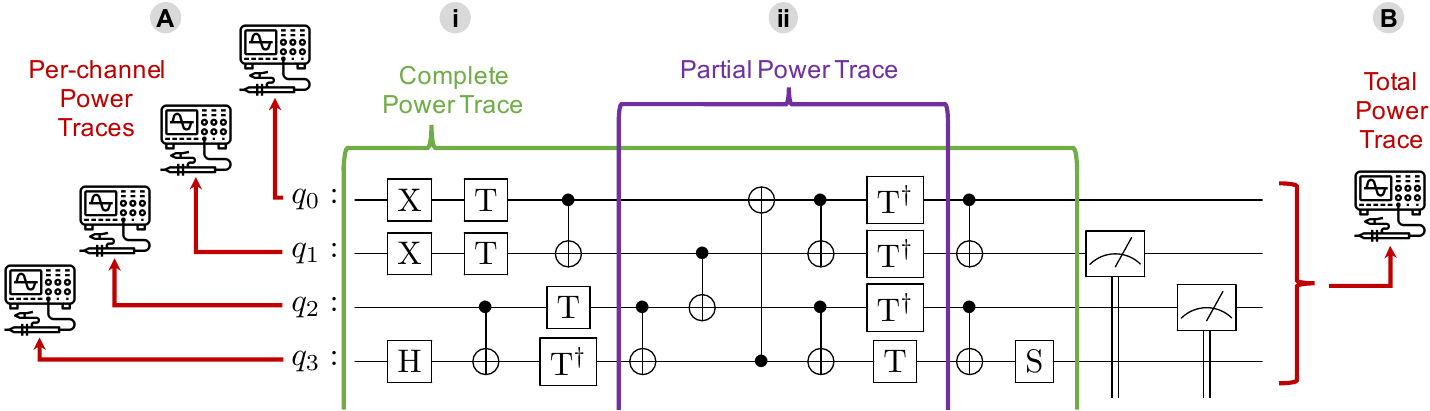}
  \caption{\small Illustration of possible measurements performed by the attacker. (A) represents per-channel power traces, while (B) represents the total power trace. Attackers can also collect data over the complete circuit, called complete power trace (i), or only a part of the circuit, called partial power trace (ii). The figure only shows which parts of the circuit the attacker targets, while the actual traces are over the pulses.}
  \label{fig_attack_taxonomy}
\end{figure*}

\subsubsection{Per-Shot Total Energy Measurements}

For a stronger attacker, we assume the attacker is able to measure the mean power and total energy of an execution of a shot of a circuit. As shown in Figure~\ref{fig_attack_taxonomy}, we assume the attacker has access to the qubit drive equipment, from which the attacker can collect the power and energy data from the arbitrary waveform generators or the mixer.

\subsubsection{Per-Shot Mean Power Measurements}

A similarly abled attacker is able to measure the mean power and total energy of an execution of a shot of a circuit. As shown in Figure~\ref{fig_attack_taxonomy}, we assume the attacker has access to the qubit drive equipment, from which the attacker can collect the power and energy data from the arbitrary waveform generators or the mixer.

\subsubsection{Per-Shot Total Power Trace Measurement}

Stronger attackers could collect a single total power trace over all channels, as shown in Figure~\ref{fig_attack_taxonomy} (B). This is more powerful than just measuring mean power or total energy.
By collecting power traces for a complete shot, shown by Figure~\ref{fig_attack_taxonomy} (i), the attacker can deploy all of our proposed attacker's goals in Sections~\ref{sec:attacker_goals}. A more powerful attacker that has knowledge of the type of circuit running, but not the oracle or the ansatz, or does not have the knowledge of the quantum processor on which the circuit ran, can measure power traces for specific portions of the shot, shown in Figure~\ref{fig_attack_taxonomy} (ii); this corresponds to our Circuit Oracle Identification (CO), Circuit Ansatz Identification (CA), and Quantum Processor Identification (QP) attacks.

\subsubsection{Per-Shot Per-Channel Power Trace Measurement}

For the strongest attacker, as shown schematically in Figure~\ref{fig_attack_taxonomy} (A), we assume the attacker is able to collect per-channel power traces. Such attackers can attempt Circuit Reconstruction (CR) attack.

\subsection{Assumptions of Attacker's Knowledge}

We want to clarify that the attacker is assumed to know at all times: the information of quantum computers (number of qubits it contains, the topology and connections of the qubits) and the basis pulse libraries of them. We assume custom gates are not used by users, and all victim circuits are composed only of the basic gates supported by the quantum computer, typically including {\tt ID}, {\tt RZ}, {\tt SX}, {\tt X}, and {\tt CX} for IBM Quantum devices. Among the basic gates, we assume the {\tt RZ} gates are virtual, as is common today. For an attacker who has only access to collect total power traces, we assume he or she knows the in-channel and cross-channel functions that define how the per-channel and total power traces correspond to the pulse information, which will be discussed in Section~\ref{sec:power_trace}.

We assume the attacker knows when the victim circuits will be executed so the attacker can capture the side-channel information. Precise knowledge of the execution time is not needed as long as the attacker can capture the trace of one shot. Since the victim often executes thousands of shots, the attacker has multiple chances to capture at least one trace. Each shot is identical without considering the noise.

\section{Experiment Setup}
\label{sec::experiment_setup}

In this paper, we used  QASMBench Benchmark Suite version 1.4~\cite{qasmbench} for NISQ evaluation.%
\footnote{We omitted benchmarks ``ipea" (iterative phase estimation algorithm) and ``shor" (Shor's algorithm) for evaluation because they have {\tt Reset} gate or in-circuit measurement that is not supported on {\tt ibm\_lagos}.}
Unless otherwise specified, {\tt ibm\_lagos}, a 7-qubit H-shape superconducting quantum computer (coupling map is shown in Figure~\ref{fig:lagos} in the appendix) is used for transpilation and scheduling. Due to the limitation of the number of qubits of {\tt ibm\_lagos}, we chose all benchmarks whose numbers of qubits are less or equal to 7. Unless otherwise specified, we used option {\tt seed\_transpiler = 0} to control the randomness and other default parameters for transpilation. Detailed information about the benchmark can be found in Table~\ref{tab:benchmarks} in the appendix.
Table~\ref{tab:benchmarks} lists details of the QASMBench benchmarks used in this~work.

\begin{table*}
  \centering
  \caption{\small QASMBench Benchmark Suite version 1.4~\cite{qasmbench}.}
  \label{tab:benchmarks}
  \small
  \begin{tabular}{llllll}
    \toprule
    {\bf Benchmark}     & {\bf Description}                             & {\bf Algorithm}        & {\bf Reference}      \\ \midrule
    deutsch             & Deutsch algorithm with 2 qubits for f(x) = x  & Hidden Subgroup        & \cite{openqasm}      \\
    iswap               & An entangling swapping gate                   & Logical Operation      & \cite{openqasm}      \\
    quantumwalks        & Quantum walks on graphs with up to 4 nodes    & Quantum Walk           & \cite{quantumwalks}  \\
    grover              & Grover's algorithm                            & Search/Optimization    & \cite{agent_anakin}  \\
    ipea*               & Iterative phase estimation algorithm          & Hidden Subgroup        & \cite{openqasm}      \\
    dnn                 & 3 layer quantum neural network sample         & Machine Learning       & \cite{dnn}           \\
    \midrule
    teleportation       & Quantum teleportation                         & Quantum Communication  & \cite{teleportation} \\
    qaoa                & Quantum approximate optimization algorithm    & Search/Optimization    & \cite{qaoa}          \\
    toffoli             & Toffoli gate                                  & Logical Operation      & \cite{scaffold}      \\
    linearsolver        & Solver for a linear equation of one qubit     & Linear Equation        & \cite{linearsolver}  \\
    fredkin             & Controlled-swap gate                          & Logical Operation      & \cite{scaffold}      \\
    wstate              & W-state preparation and assessment            & Logical Operation      & \cite{openqasm}      \\
    basis\_change       & Transform the single-particle basis           & Quantum Simulation     & \cite{openfermion}   \\
    \midrule
    qrng                & Quantum random number generator               & Quantum Arithmetic     & \cite{qrng}          \\
    cat\_state          & Coherent superposition of two coherent states & Logical Operation      & \cite{scaffold}      \\
    inverseqft          & exact inversion of quantum Fourier tranform   & Hidden Subgroup        & \cite{openqasm}      \\
    adder               & Quantum ripple-carry adder                    & Quantum Arithmetic     & \cite{scaffold}      \\
    hs4                 & Hidden subgroup problem                       & Hidden Subgroup        & \cite{scaffold}      \\
    bell                & Circuit equivalent to Bell inequality test    & Logic Operation        & \cite{cirq}          \\
    qft                 & Quantum Fourier transform                     & Hidden Subgroupe       & \cite{openqasm}      \\
    variational         & Variational ansatz for a Jellium Hamiltonian  & Quantum Simulation     & \cite{openfermion}   \\
    vqe\_uccsd          & Variational quantum eigensolver with UCCSD    & Linear Equation        & \cite{scaffold}      \\
    basis\_trotter      & Trotter steps for molecule LiH at equilibrium & Quantum Simulation     & \cite{openfermion}   \\
    \midrule
    qec\_sm             & Repetition code syndrome measurement          & Error Correction       & \cite{openqasm}      \\
    lpn                 & Learning parity with noise                    & Machine Learning       & \cite{sampaio96}     \\
    qec\_en             & Quantum repetition code encoder               & Error Correction       & \cite{sampaio96}     \\
    shor*               & Shor's algorithm                              & Hidden Subgroup        & \cite{ibm_qiskit}    \\
    pea                 & Phase estimation algorithm                    & Hidden Subgroup        & \cite{openqasm}      \\
    error\_correctiond3 & Error correction with distance 3 and 5 qubits & Error Correction       & \cite{bench_2017}    \\
    \midrule
    simons              & Simon's algorithm                             & Hidden Subgroup        & \cite{agent_anakin}  \\
    qaoa                & Quantum approximate optimization algorithm    & Search \& Optimization & \cite{cirq}          \\
    vqe\_uccsd          & Variational quantum eigensolver with UCCSD    & Linear Equation        & \cite{scaffold}      \\
    \midrule
    hhl                 & HHL algorithm to solve linear equations       & Linear Equation        & \cite{ibm_hhl}       \\
    \bottomrule
  \end{tabular}\\
  \footnotesize{* These circuits contain the middle measurement and {\tt Reset} gate, and cannot be scheduled on the backend currently because their basis pulses are not provided.}\\
\end{table*}

\subsection{Power Traces}
\label{sec:power_trace}

In the experiments, the total power traces, the per-channel power traces, and the pulse amplitude time series are all one-dimensional time series. For the in-channel and across-channel functions discussed in Section~\ref{sec:threat_model_power_trace}, we assume:
\begin{equation}
    Power_c[p_c(x)] = \text{Re}^2[p_c(x)] + \text{Im}^2[p_c(x)]
\end{equation}
and:
\begin{equation}
    Total[f_{c_1}(x), \dots, f_{c_n}(x)] = \sum_{i \in \{1, \dots, n\}} f_{c_i}(x)
\end{equation}
which means the per-channel power traces are the square of the norm of the amplitude, and the total power traces are directly the summation of per-channel power traces with the same weight. 

In our experiments, we obtained the pulse information from Qiskit APIs provided by IBM Quantum on each of the target quantum computers. From the pulse information, we computed the per-channel and the total power traces using the above functions.

\subsection{Circuit Norm and Distance}
\label{sec:dist}

In the evaluation, we define 3 metrics: {\em circuit norm}, {\em circuit distance} between two circuits, and {\em normalized circuit distance} between two circuits, all of which are in terms of the total power traces:

\begin{enumerate}
    \item $norm(C)$: the circuit norm of the circuit $C$ with the total power traces $P_C(x)$ is $f_{norm}[P_C(x)]$
        
    \item $dist(C_1, C_2)$: the circuit distance of the circuit $C_1$ and the circuit $C_2$ is $f_{dist}[P_{C_1}(x), P_{C_2}(x)]$.
        
    \item $norm\_dist(C_1, C_2)$: the normalized circuit distance of the circuit $C_1$ and the circuit $C_2$ is $\frac{1}{norm(C_1)} dist(C_1, C_2)$. 
\end{enumerate}
For attackers, a bigger circuit distance between circuit $C_1$ and $C_2$ means it is easier to identify these two circuits. The definitions depend on the choice of the norm $f_{norm}$ and distance function $f_{dist}$. In this paper, we choose the Euclidean norm and distance for these two functions, i.e., $f_{norm}(\vec a) = \sqrt{\sum_{i=1}^n a_i^2}$ and $f_{dist}(\vec a, \vec b) = \sqrt{\sum_{i=1}^n (a_i - b_i)^2}$.

\section{Attack Evaluation}
\label{sec_evaluation}

In this section, we evaluate all the attackers' goals listed in Section~\ref{sec:attacker_goals}. 

\subsection{User Circuit Identification (UC)}
\label{sec:uc}

For UC evaluation, we started with the QASMBench benchmarks. To further expand the circuit list, we chose different initial layouts in the transpilation so that the same circuit can be transpiled into different circuits based on the hardware configuration. For an $n$-qubit circuit on $k$-qubit backend, the number of initial layouts is in total ${n \choose k}$. In the experiment, we chose 8 circuit lists $CL_{i}$, where $i$ is the number of initial layouts. We choose $i$ to be 1, 2, 4, 8, 16, 32, 64, 128. The exact initial layout is randomly selected from ${n \choose 7}$ initial layouts. If for one circuit, $i > {n \choose 7}$, which means there are not enough initial layouts, then we choose all the ${n \choose 7}$ permutations as the initial layouts. For reference, after expanding, the number of circuits in the circuit list is listed in Table~\ref{table:no_circ}.

Besides the total power traces, three additional metrics are also used to evaluate the results: energy, mean power, and duration of the circuit. The energy is computed by adding all terms of the one-dimensional total power time series, which is the total energy in the {\tt dt} unit of the circuit. The duration is the time from the start to the end of the circuit in the {\tt dt} time unit%
\footnote{$1$dt $=0.222$ns, which is a time unit used in IBM Quantum.}, which is also the same as the length of the one-dimensional total power time series. The mean power is then computed by dividing the energy by the duration. For a circuit $C$, we used $m_p(C)$, $m_e(C)$, $m_m(C)$, and $m_d(C)$ to represent these values.

For the UC experiment (i.e. identifying the user circuit from a known list of circuits), we define the accuracy to be the proportion of circuits in the circuit list that are correctly identified. More specifically, for each circuit $C \in CL_{i}$, we calculated the distance $dist(x)$ (see Section~\ref{sec:dist}) with the metric $m(x)$ between it and all the circuits in the list:
\begin{equation}
  dist[m(C), m(C^\prime)],\ \forall C^\prime \in CL_{i}
\end{equation}
The identification for the circuit $C$ is chosen to be the circuit with the smallest distance between the measured and the software-generated metric of this circuit:
\begin{equation}
  \text{id}_{i, m}(C) = \argmin_{C^\prime \in CL_{i}} dist[m(C), m(C^\prime)]
\end{equation}

In addition, we simulated the potential practical environment of gathering leaked information. The measurement error $e(x)$ was introduced when computing the metrics. With the error, the presumptive measured metric is added by the error, while other metrics are software-generated and not influenced by the error. Specifically, with error $e(x)$, the identification is changed to:
\begin{equation}
  \text{id}_{i, m, e}(C) = \argmin_{C^\prime \in CL_{i}} dist[m_e(C), m(C^\prime)]
\end{equation}
where
\begin{equation}
  m_e(C) = m(C) + e[m(C)]
\end{equation}
in the experiment, the error has the same length as the metric. The error value is randomly chosen from the normal distribution with the expectation to be 0 and the standard deviation to be the error rate, and then multiplied by the metric value.

\begin{table}[]
  \caption{\small Number of possible layouts and the corresponding number of circuits used in user circuit identification (UC) experiments.}
  \label{table:no_circ}
  \small
  \hspace{-1.5em}
  \begin{tabular}{|c|c|c|c|c|c|c|c|c|}
    \hline
    \textbf{No. Layouts}  & 1  & 2  & 4   & 8   & 16  & 32  & 64   & 128  \\ \hline
    \textbf{No. Circuits} & 31 & 62 & 124 & 248 & 496 & 992 & 1874 & 3538 \\ \hline
  \end{tabular}
\end{table}

\begin{figure*}[t]
  \captionsetup[subfigure]{justification=centering}
  \centering
  \begin{subfigure}[t]{0.27\textwidth}
    \centering
    \includegraphics[width=\textwidth]{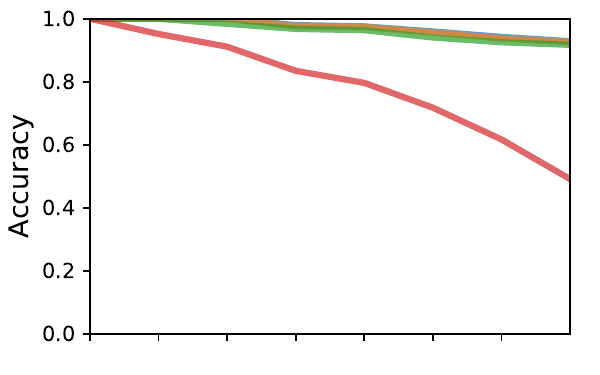}
    \caption{Error Rate: 0.0}
  \end{subfigure}%
  ~
  \begin{subfigure}[t]{0.27\textwidth}
    \centering
    \includegraphics[width=\textwidth]{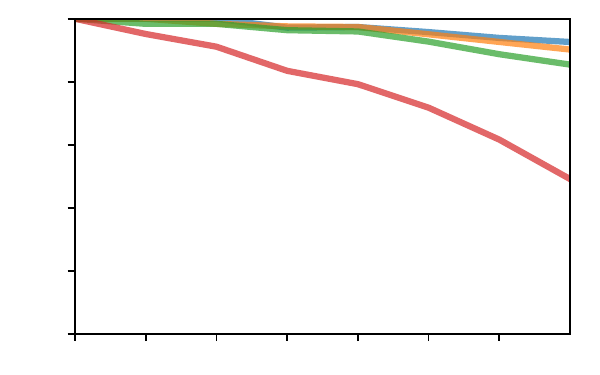}
    \caption{Error Rate: 0.00001}
  \end{subfigure}%
  ~
  \begin{subfigure}[t]{0.27\textwidth}
    \centering
    \includegraphics[width=\textwidth]{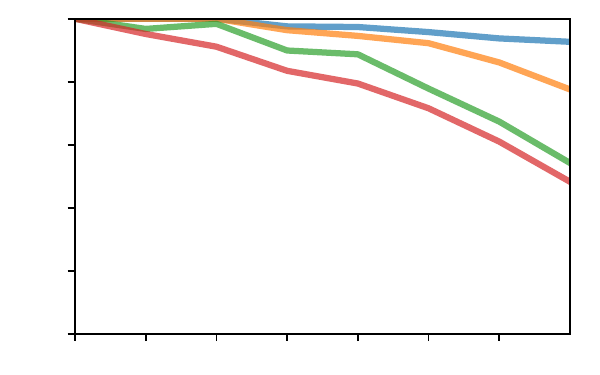}
    \caption{Error Rate: 0.0001}
  \end{subfigure}%
  \\
  \begin{subfigure}[t]{0.27\textwidth}
    \centering
    \includegraphics[width=\textwidth]{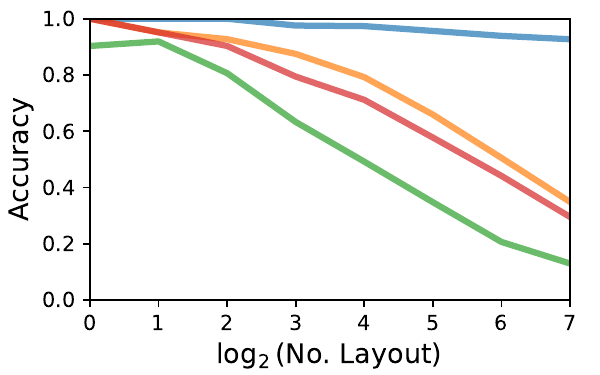}
    \caption{Error Rate: 0.001}
  \end{subfigure}%
  ~
  \begin{subfigure}[t]{0.27\textwidth}
    \centering
    \includegraphics[width=\textwidth]{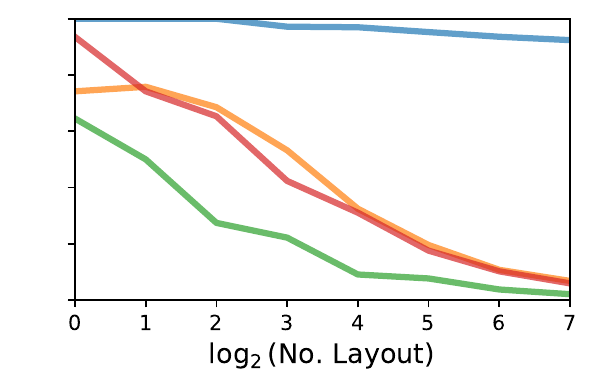}
    \caption{Error Rate: 0.01}
  \end{subfigure}%
  ~
  \begin{subfigure}[t]{0.27\textwidth}
    \centering
    \includegraphics[width=\textwidth]{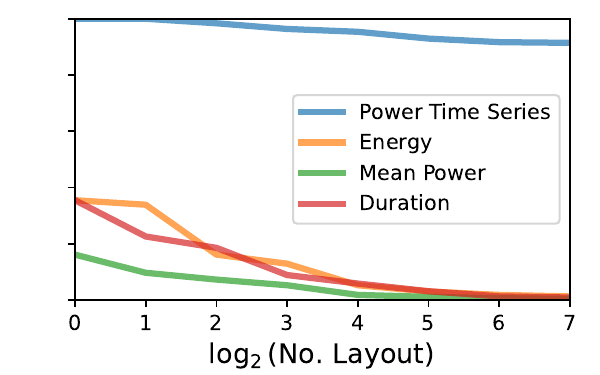}
    \caption{Error Rate: 0.1}
  \end{subfigure}%
  \caption{\small Evaluation for user circuit identification (UC). Accuracy based on 4 metrics: power time series, energy, mean power, and duration. The circuit set is made up of QASMBench and expanded by transpiling with a number of initial layouts. Errors are simulated by randomly sampling from the normal distribution whose expected value is 0, the standard deviation is the error rate, and the length is the same as the metrics. The measured metrics are then added by the error times themselves.}
  \label{fig:accuracy}
\end{figure*}

Figure~\ref{fig:bm_energy} -- Figure~\ref{fig:bm_duration} shows the energy, mean power, and duration of the original benchmark. The figure is shown later in the paper as it also includes the same metrics when our defenses are applied. The distribution of the metrics' values gives an insight into how these physical quantities perform in identifying user circuits. Based on the experiment setup above, we computed the accuracy, which is shown in Figure~\ref{fig:accuracy}. As the figure shows, though power-related traces are harder to gather than timing traces, they have a better performance when identifying user circuits. As the number of layouts increases, the accuracy computing by duration decreases much more than power-related metrics. One reason is that duration is in {\tt dt} unit, making it easier to be the same for different circuits, while power-related metrics are more distinct from each other. 

\begin{figure*}
  \captionsetup[subfigure]{justification=centering}
  \centering
  \begin{subfigure}[t]{0.32\textwidth}
    \centering
    \includegraphics[width=0.7\textwidth]{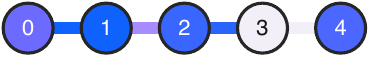}
    \caption{ibmq\_manila (Falcon r5.11L)\\Basis Gates: \{CX, ID, RZ, SX, X\}}
    \label{fig:manila}
  \end{subfigure}
  ~
  \begin{subfigure}[t]{0.32\textwidth}
    \centering
    \includegraphics[width=0.4\textwidth]{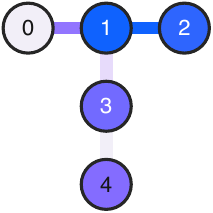}
    \caption{ibmq\_lima (Falcon r4T)\\Basis Gates: \{CX, ID, RZ, SX, X\}}
    \label{fig:lima}
  \end{subfigure}%
  \begin{subfigure}[t]{0.32\textwidth}
    \centering
    \includegraphics[width=0.4\textwidth]{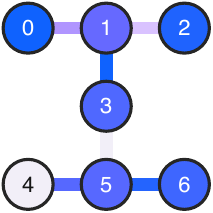}
    \caption{ibm\_lagos (Falcon r5.11H)\\Basis Gates: \{CX, ID, RZ, SX, X\}}
    \label{fig:lagos}
  \end{subfigure}

  \begin{subfigure}[t]{0.33\textwidth}
    \centering
    \hspace{-1cm}
    \includegraphics[width=\textwidth]{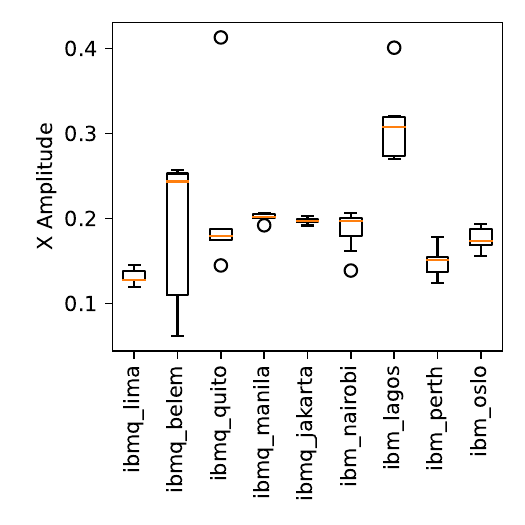}
    \caption{Amplitude of {\tt X}}
    \label{fig:amp_x}
  \end{subfigure}%
  ~
  \begin{subfigure}[t]{0.33\textwidth}
    \centering
    \hspace{-1cm}
    \includegraphics[width=\textwidth]{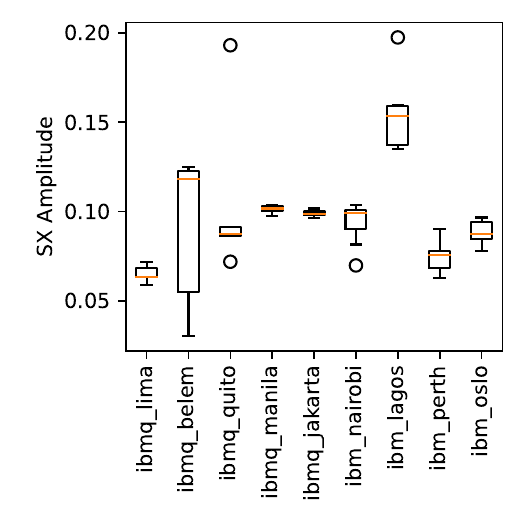}
    \caption{Amplitude of {\tt SX}}
    \label{fig:amp_sx}
  \end{subfigure}%
  ~
  \begin{subfigure}[t]{0.33\textwidth}
    \centering
    \hspace{-1cm}
    \includegraphics[width=\textwidth]{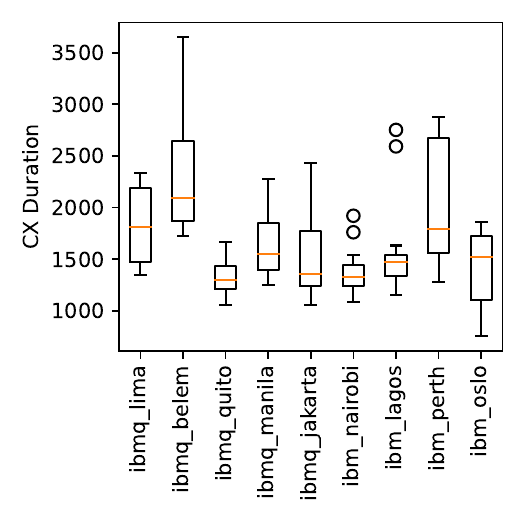}
    \caption{Duration of {\tt CX}}
    \label{fig:duration_cx}
  \end{subfigure}
  \caption{IBM Quantum device information. (a) -- (c) Three coupling maps of the IBM-Q devices. The color of nodes implies the frequency (GHz) of the qubit (GHz, darker color means lower frequency). The connection color implies the gate time in nanoseconds for 2-qubit gates such as CX (darker color means shorter time). (d) -- (e) Box plots of amplitude of {\tt X} and {\tt SX} and duration of {\tt CX} on 9 IBM Quantum backends. }
  \label{fig:backend_info}
\end{figure*}

We also consider the case of noise or errors in the side-channel information. Firstly, the accuracy based on the power time series is much more stable over different error rates and thus has a better distinguishability than the other three metrics. One reason is that the power time series is a one-dimensional array, while the other three metrics are only scalars. Therefore, it needs much larger noise for the attacker to make a wrong identification based on the power time series.

Secondly, with small error rates, power-related metrics (power time series, energy, and mean power) have better performance than the duration, while with large error rates, the duration is better than the mean power, but is similar to the energy. One reason is that the distribution of the mean power for quantum circuits is more centralized than the distribution of the duration, which is also shown in Figure~\ref{fig:bm_mean_power} and \ref{fig:bm_duration}, since the mean power is the average over the power on all the time steps. Also, the duration of quantum circuits can be arbitrary, while the upper bound of the mean power is limited by the summation of the native gates with the largest mean power. On the other hand, energy encodes both the duration of quantum circuits and information about the gates and quantum hardware. The choice between using the energy or the duration as the metric may depend on the use cases. In the case that quantum circuits in the circuit list have similar duration, the energy can perform better than the duration. On the other hand, in the case that quantum circuits have similar energies, the duration is a better metric for attackers to collect.

\subsubsection*{UC Attack Summary:}
Timing, total energy, and mean power attacks are able to identify user circuits with very high accuracy, reaching close to $100$\% when attackers have zero or very small errors in the side-channel information. Timing attacks perform worse than total energy and mean power with a small noise, while it is similar to energy and better than mean power with a large noise. Meanwhile, power trace attacks are always the best and robust over different noise levels.

\subsection{Circuit Oracle Identification (CO)}
\label{sec:co}

Many quantum algorithms consist of oracles, which act like black boxes that return desired quantum states based on the input. For example, a {\em Boolean oracle} changes the input states to another binary representation, i.e., $U_f \ket x \otimes \ket{\bar 0} = \ket x \otimes \ket{f(x)}$; a {\em phase oracle} does not change the state but change its phase, i.e., $P_f \ket x = (-1)^{f(x)} \ket x$.

For CO, we choose three textbook algorithms for evaluating how the oracle can be identified with the quantum computing power side-channels:
\begin{enumerate}
  \item Bernstein-Vazirani (BV)~\cite{doi:10.1137/S0097539796300921}: given an oracle $f(x) = s \cdot x$, find the hidden $s$ in the oracle.
  \item Deutsch-Jozsa (DJ)~\cite{deutsch1992rapid}: given an oracle $f(x) = 0\ \text{or}\ 1$, which is either a constant function whose outputs are all 0 or all 1, or a balanced function whose outputs are half 0 and half 1, find whether the oracle is constant or balanced.
  \item Grover’s Search (GS)~\cite{10.1145/237814.237866}: given an oracle $f(x)$ to reflect the states, find a state specified by the oracle.
\end{enumerate}

All these algorithms can have an arbitrary number of qubits. We tested from 1-qubit to 6-qubit versions, and for all the $n$-qubit algorithms, the parameters specifying the oracles are tested from $0\cdots 0$ to $1\cdots 1$. Since if the function for DJ is constant, the oracle can be an empty circuit, we only tested the balanced function.

\begin{table}[]
  \centering
  \caption{\small Evaluation for circuit oracle identification (CO). Normalized circuit distance for Bernstein-Vazirani, Deutsch-Jozsa, and Grover's Search with the number of qubits from 1 to 6 on {\tt ibm\_lagos}. Bernstein-Vazirani and Deutsch-Jozsa need one additional qubit to control the oracle. Bigger value means oracles can be more easily identified.}
  \label{tab:oracle}
  \small
  \begin{tabular}{|c|cccccc|}
    \hline
    \multirow{2}{*}{\textbf{Algorithm}} & \multicolumn{6}{c|}{\textbf{Number of Qubits/Oracles}}                                                                                                                                                                   \\ \cline{2-7}
                                        & \multicolumn{1}{c|}{\textbf{1/2}}                      & \multicolumn{1}{c|}{\textbf{2/4}} & \multicolumn{1}{c|}{\textbf{3/8}} & \multicolumn{1}{c|}{\textbf{4/16}} & \multicolumn{1}{c|}{\textbf{5/32}} & \textbf{6/64} \\ \hline
    Bernstein-Vazirani                  & \multicolumn{1}{c|}{1.00}                              & \multicolumn{1}{c|}{0.30}         & \multicolumn{1}{c|}{0.07}         & \multicolumn{1}{c|}{0.06}          & \multicolumn{1}{c|}{0.07}          & 0.06          \\ \hline
    Deutsch-Jozsa                       & \multicolumn{1}{c|}{0.00}                              & \multicolumn{1}{c|}{0.00}         & \multicolumn{1}{c|}{0.00}         & \multicolumn{1}{c|}{0.00}          & \multicolumn{1}{c|}{0.00}          & 0.00          \\ \hline
    Grover's Search                     & \multicolumn{1}{c|}{0.00}                              & \multicolumn{1}{c|}{0.00}         & \multicolumn{1}{c|}{0.00}         & \multicolumn{1}{c|}{0.00}          & \multicolumn{1}{c|}{0.00}          & 0.00          \\ \hline
  \end{tabular}
\end{table}

\begin{figure*}
    \centering
     \begin{subfigure}[b]{\textwidth}
         \centering
         \input{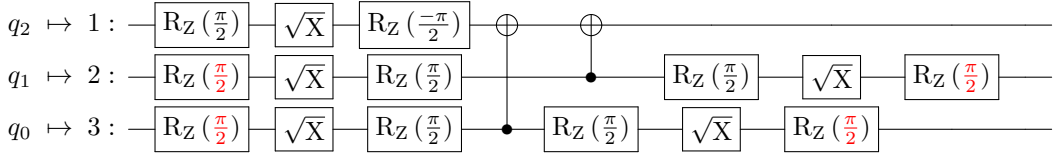}
         \caption{\small 3-qubit Deutsch-Jozsa with 00 as the hidden string.}
         \label{fig:dj_circuit}
     \end{subfigure}

     \begin{subfigure}[b]{\textwidth}
         \centering
         \input{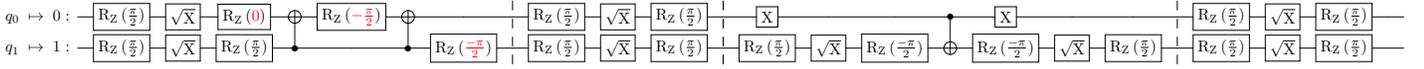}
         \caption{\small 2-qubit Grover's search with $|00\rangle$ as the target state.}
         \label{fig:gs_circuit}
     \end{subfigure}

    \begin{subfigure}[b]{\textwidth}
         \centering
         \input{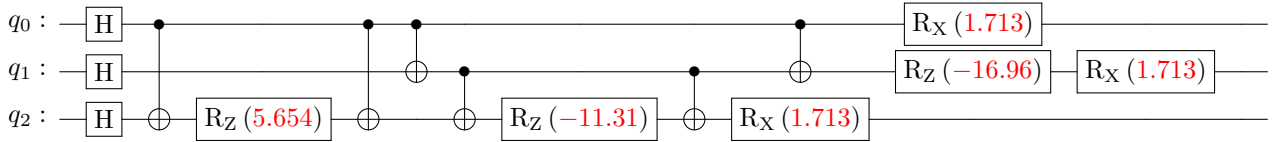}
         \caption{\small 3-qubit quantum approximate optimization algorithm.}
         \label{fig:qaoa}
     \end{subfigure}
     
    \caption{\small Schematic of quantum circuits. (a)-(b) Quantum circuits for Deutsch-Jozsa and Grover's Search. The red color shows the possible changes in the angles of {\tt RZ} gates to realize different oracles. (c) The quantum circuit for the quantum approximate optimization algorithm. The red color shows the possible changes in the optimization process. Global phases are not neglected in figures.}
    \label{fig:circ_plot}
\end{figure*}

The minimum normalized circuit distance is used to evaluate the results, shown in Table~\ref{tab:oracle}. For BV, since the oracles are quite different from each other, the minimum circuit distance is not 0, which means the oracles can be distinguished from each other. However, for DJ and GS, the circuits for different oracles can be the same, and the only changes are the angles of the rotation gates, such as {\tt RZ} gate. As an example, we show in Figure~\ref{fig:dj_circuit} and \ref{fig:gs_circuit} when appropriately changing the angles in red color, the oracle can be changed. Since {\tt RZ} is a virtual gate on IBM quantum backends with no duration and amplitudes, all circuits have the same power traces and thus cannot be distinguished from each other. More details of the virtual {\tt RZ} gate will be discussed in Section~\ref{sec:defense_rz}.

Another thing that needs to pay attention to for circuit oracle identification is that circuits after transpilation are highly dependent on the transpiler settings. For example, the oracles of some algorithms have symmetries, such as 3-qubit Bernstein-Vazirani with "01" and "10" as the hidden string, the transpiler may output the same circuits. This can be achieved by changing the bit order of the measurement results. 

\subsubsection*{CO Attack Summary:}
Whether quantum computer power side-channels can be exploited to retrieve the information of oracles depend on the algorithm. Oracles changing the gate types can be easily distinguished, while oracles only changing the rotational angles in the virtual {\tt RZ} gates are hard to distinguish.

\subsection{Circuit Ansatz Identification (CA)}

One important application of quantum computing is solving optimization problems, such as finding the minimum eigenvalue of a matrix. The Variational Quantum Eigensolver (VQE)~\cite{peruzzo2014variational} and the Quantum Approximate Optimization Algorithm (QAOA)~\cite{farhi2014quantum} are the representative quantum algorithms for optimization. Besides, quantum machine learning~\cite{biamonte2017quantum} and quantum deep learning~\cite{wiebe2014quantum} are also actively researched algorithms. These algorithms solve the optimization problem by generating appropriate quantum states through parameterized circuits and iteratively updating parameters to find the extremes. These circuits are also often called~{\em ansatz}. Finding out what the ansatz is can enable attackers to, for example, learn the types of algorithms used by the victim.

For demonstrating ability to identify circuit ansatz, we chose $6$ ansatz circuits from the benchmarks "qaoa\_n3", "variational\_n4", "vqe\_n4", "vqe\_uccsd\_n4", "qaoa\_n6", and "vqe\_uccsd\_n6", and computed the minimum normalized circuit distance between these circuits, which is $0.97$. Such a large normalized circuit distance proves the ability to effectively distinguish them.

In addition to the ansatz circuit configuration, another important piece of information about the ansatz circuit is its parameters, such as red values highlighted in Figure~\ref{fig:qaoa} in the appendix. However, due to the same reason discussed in Section~\ref{sec:co} why oracle for Deutsch-Jozsa or Grover's search cannot be identified, the parameters usually only change the rotational angles of the virtual {\tt RZ} gates in the ansatz circuit, while other real gates remain the same, it is impossible to retrieve any information from the power traces about the parameters. More discussion about the virtual {\tt RZ} gate will be discussed in Section~\ref{sec:defense_rz}. 

\subsubsection*{CA Attack Summary:}

Attackers can identify which ansatz was used, but the parameters of the ansatz cannot be easily recovered by the attackers. Frequent use of virtual {\tt RZ} gates in the ansatz makes them naturally less vulnerable to attacks.

\subsection{Qubit Mapping Identification (QM)}

As discussed in previous sections, the pulses for one quantum gate on different qubit or qubit pairs are different since the pulses need to be calibrated based on the qubit's physical properties to achieve the same logical operations. Thus, the power traces also encode the information of the physical qubits to which the quantum gates are applied to.

Before the quantum circuit is executed on the quantum device, the mapping from the logical qubits to the physical qubits must be specified. In the transpilation process of Qiskit, the qubit mapping is automatically selected if no input for the layout is given. In the experiment, we selected $10$ initial layouts for each circuit in the benchmark, and compute the minimum normalized circuit distance in the circuit list.

The results are shown in the QM column of Table~\ref{tab:eval_bench}. Nearly all of the benchmarks have a large minimum normalized circuit distance, which indicates that they can be well distinguished from each other. However, the minimum normalized circuit distance of ``inverseqft" (inverse quantum Fourier transformation) and ``qrng" (quantum random number generator) is 0. The reason is that the circuits for both these algorithms only consist of single-qubit gates (``inverseqft" also has the dynamical {\tt RZ} gate), so when changing the order of the qubits in the initial layout, it does nothing to the circuit. For example, the circuits with initial layout [0, 1, 2, 3] and [1, 0, 2, 3] are the same, and therefore the circuit distance is 0 between these two circuits with such initial layouts. However, the circuit distance is not 0 if the initial layouts contain at least 1 different qubit. 

\begin{figure}
  \centering
  \includegraphics[width=0.5\textwidth]{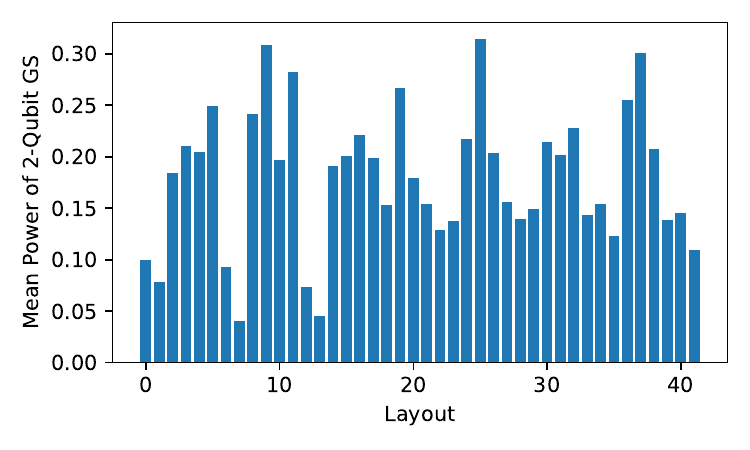}
  \caption{One example of mean power with different layouts of 2-qubit Grover's search showing that the power traces are also highly dependent on hardware.}
  \label{fig:grover_mean_power}
\end{figure}

\subsubsection*{QM Attack Summary:}

For most circuits, attackers are able to determine from the power traces what was the assignment of physical qubits to the qubits in the circuit, making this a feasible attack.

\begin{table}[]
\caption{\small Evaluation for qubit mapping (QM) identification, quantum processor (QP) identification, and circuit reconstruction (CR). The benchmark parameters, such as numbers of gates, are based on circuits transpiled on {\tt ibm\_lagos} with {\tt seed\_transpiler = 0} and other default arguments. The minimum normalized circuit distance is used to evaluate the results for QM and QP. For RP, the checkmark shows the non-virtual gates in the original circuit are correctly reconstructed given the per-channel power traces.}
\label{tab:eval_bench}
\small
\begin{tabular}{l|lll|lll}
\toprule
\textbf{QASMBench} & \multicolumn{3}{c|}{\textbf{Parameters}} & \multicolumn{3}{c}{\textbf{Attacks}} \\
\textbf{Benchmark} & \textbf{Qubit} & \textbf{Gate} & \textbf{CX} & \textbf{QM} & \textbf{QP} & \textbf{CR}\\
\midrule
deutsch            & 2              & 10            & 1           & 0.025       & 0.116       & \checkmark   \\
dnn                & 2              & 306           & 42          & 0.039       & 0.116       & \checkmark  \\
grover             & 2              & 15            & 2           & 0.143       & 0.116       & \checkmark    \\
iswap              & 2              & 14            & 2           & 0.143       & 0.116       & \checkmark      \\
quantumwalks       & 2              & 38            & 3           & 0.125       & 0.117       & \checkmark   \\ \midrule
basis\_change      & 3              & 85            & 10          & 0.673       & 0.068       & \checkmark    \\
fredkin            & 3              & 31            & 17          & 0.800       & 0.411       & \checkmark      \\
linearsolver       & 3              & 26            & 4           & 0.735       & 0.080       & \checkmark       \\
qaoa               & 3              & 35            & 9           & 0.546       & 0.570       & \checkmark     \\
teleportation      & 3              & 12            & 2           & 0.473       & 0.075       & \checkmark     \\
toffoli            & 3              & 24            & 9           & 0.096       & 0.573       & \checkmark      \\
wstate             & 3              & 47            & 21          & 0.789       & 0.101       & \checkmark     \\ \midrule
adder              & 4              & 33            & 16          & 0.727       & 0.201       & \checkmark      \\
basis\_trotter     & 4              & 2353          & 582         & 0.895       & 0.220       & \checkmark       \\
bell               & 4              & 53            & 7           & 0.781       & 0.196       & \checkmark     \\
cat\_state         & 4              & 6             & 3           & 0.744       & 0.241       & \checkmark      \\
hs4                & 4              & 28            & 4           & 0.545       & 0.327       & \checkmark      \\
inverseqft         & 4              & 30            & 0           & 0.000       & 0.001       & \checkmark      \\
qft                & 4              & 50            & 18          & 0.817       & 0.287       & \checkmark     \\
qrng               & 4              & 12            & 0           & 0.000       & 0.001       & \checkmark      \\
variational        & 4              & 58            & 16          & 0.792       & 0.239       & \checkmark       \\
vqe                & 4              & 73            & 9           & 0.660       & 0.194       & \checkmark   \\
vqe\_uccsd         & 4              & 238           & 88          & 0.858       & 0.241       & \checkmark    \\ \midrule
error\_c3          & 5              & 249           & 61          & 0.855       & 0.220       & \checkmark      \\
lpn                & 5              & 17            & 2           & 0.576       & 0.194       & \checkmark       \\
pea                & 5              & 126           & 57          & 0.874       & 0.210       & \checkmark     \\
qec\_en            & 5              & 52            & 16          & 0.746       & 0.250       & \checkmark     \\
qec\_sm            & 5              & 8             & 4           & 0.573       & 0.266       & \checkmark    \\ \midrule
qaoa               & 6              & 408           & 84          & 0.869       & 0.283       & \checkmark      \\
simon              & 6              & 65            & 23          & 0.796       & 0.605       & \checkmark      \\
vqe\_uccsd         & 6              & 2289          & 1199        & 0.906       & 0.278       & \checkmark     \\ \midrule
hhl                & 7              & 1092          & 298         & 0.873       & 0.317       & \checkmark    \\ \bottomrule
\end{tabular}
\end{table}

\subsection{Quantum Processor Identification (QP)}

Another kind of hardware-related information can be the quantum processor on which the circuit was executed. 
The identification among quantum processors with distinct connections may be easier for circuits with a large number of qubits since it needs to add switch gates to the circuit and the information of quantum processors is encoded in terms of connections. Nevertheless, the identification among quantum processors with the same coupling map is also feasible since the properties of qubits are distinct across quantum processors and this information is included in the basis pulse library.

We selected 9 IBM Quantum backends to show the diversity among quantum devices: {\tt ibmq\_lima}, {\tt ibmq\_quito}, {\tt ibmq\_belem}, {\tt ibmq\_manila}, {\tt ibmq\_jakarta}, {\tt ibm\_oslo}, {\tt ibm\_nairobi}, {\tt ibm\_lagos}, {\tt ibm\_perth}. The former 4 devices are 5-qubit and the others are 7-qubit devices. There are two coupling maps for 5-qubit devices: line-shape shown in Figure~\ref{fig:manila} and T-shape shown in Figure~\ref{fig:lima}, and only one coupling map for the 7-qubit devices: H-shape shown in Figure~\ref{fig:lagos}. The statistics of the amplitude of {\tt X} and {\tt SX} gates on different qubits are shown in Figure~\ref{fig:amp_x} and Figure~\ref{fig:amp_sx}, and the statistics of the duration of {\tt CX} gates is shown in Figure~\ref{fig:duration_cx}. The figures are in the appendix. All of them have distinct features in the basis pulse library. Note that the distribution of {\tt X} and {\tt SX} are the same. This is due to that only {\tt X} is calibrated, and the amplitude of {\tt SX} is directly set to be half of the amplitude of {\tt X}. 

To quantify the influence of the difference of the connectivity and basis pulse library over backends on the total power traces of quantum circuits, we transpiled the benchmark on these 9 quantum devices. The QP column of Table~\ref{tab:eval_bench} shows the minimum normalized circuit distance over these devices. Most of the circuits have large enough circuit distances over different quantum devices, making them straightforward to be separated individually. In addition, ``inverseqft" and ``qrng" may not be determined for qubit mapping identification, but they are possible to be recognized for quantum processor identification. 

\subsubsection*{QP Attack Summary:}

For most circuits, attackers are able to correctly identify on which backend they were executed, making this a feasible attack.

\subsection{Circuit Reconstruction (CR)}
\label{sec:rp}

The most powerful attacker we analyze is one who has access to per-channel power traces.
We implement an algorithm to reconstruct the circuit and the results are shown in the CR column of Table~\ref{tab:eval_bench}. We can successfully reconstruct all circuits in the benchmark given their per-channel power traces.

\begin{figure}[t]
  \centering
  \includegraphics[width=1.0\columnwidth]{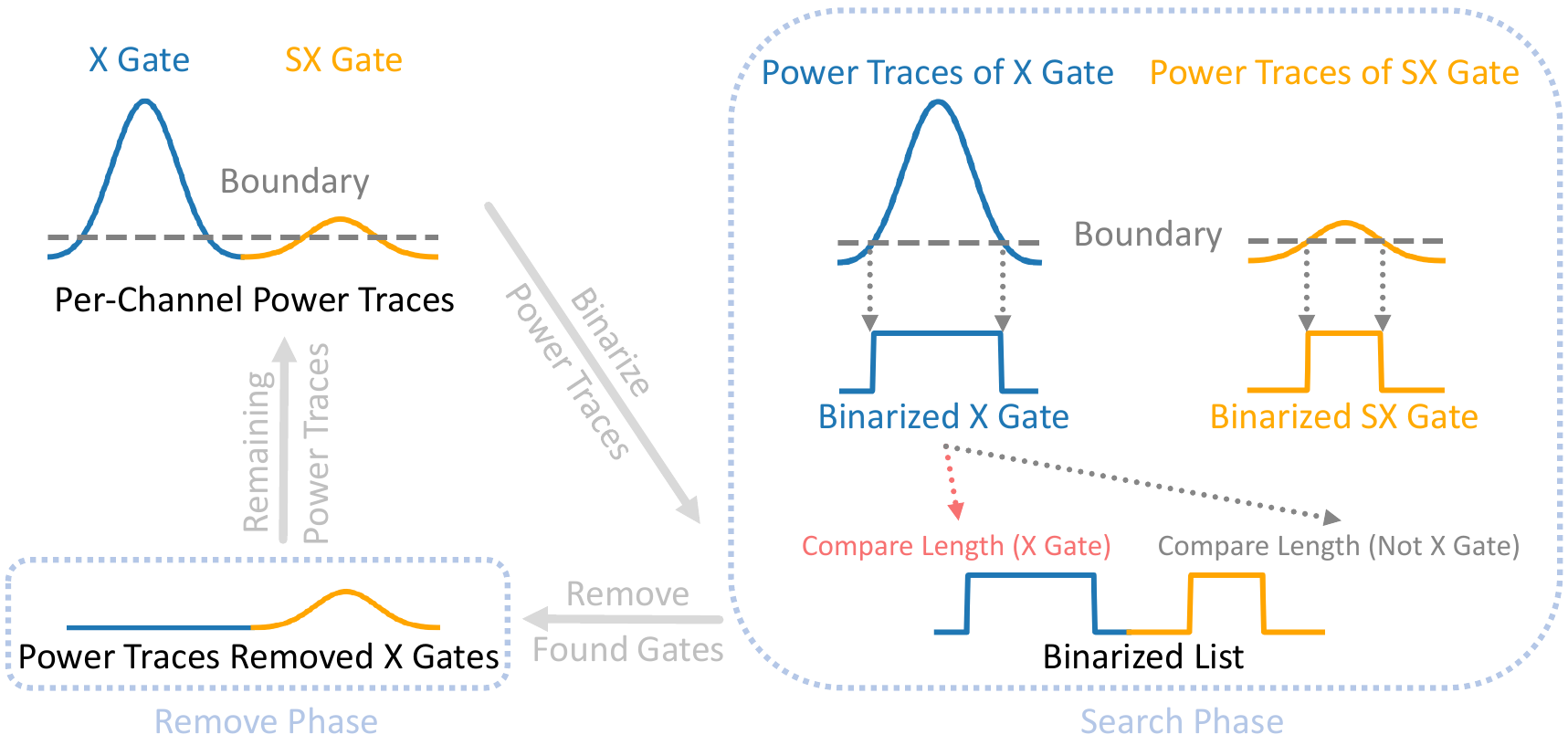}
  \caption{\small Algorithm for circuit reconstruction. The algorithm includes two phases: the search phase and the remove phase. In the search phase, the algorithm binarizes the power traces and searches for a target gate in the power traces by comparing the length of the binary segments with the length of the binarized power traces of the basis gates. In the remove phase, the algorithm removes all the target gates from the power traces and generates the new power traces for the next iteration.}
  \label{fig:rp_algo}
\end{figure}

The algorithm is illustrated in Figure~\ref{fig:rp_algo}. The algorithm iterates over all channels and finds the corresponding pulses. The algorithm includes two phases: the {\em search phase} and {\em remove phase}. In the search phase, the algorithm locates all gates in the power traces and selects the target gate. In the remove phase, the algorithm removes all the target gates from the power traces and generates new power traces without the removed gates for the next iteration.

While multi-qubit gates may include several pulses on several channels, and some of these pulses may have the same shape as the single-qubit pulses, our implementation first iterates all control channels and find all multi-qubit gates. After locating all multi-qubit gates, the algorithm removes them from the per-channel power traces. Then a similar process is done for single-qubit gates. The algorithm iterates the remaining drive channels and locates specific single qubit gates, and then removes them from the per-channel power traces. After iterating all channels and all basis gates, the found gates and their start times are the output of the algorithm.

For IBM Quantum backends, there are only three real gates, {\tt X}, {\tt SX}, and {\tt CX}. 
We transform the goal of finding the pulses (representing the gates) in the power traces into finding the segment in the binary list. This is done by binarizing the per-channel traces based on an input boundary, i.e., if the power is larger than the boundary, its value is set to be 1, and set to 0 if not. The same process is also done for the software-generated power traces of basis gates. After binarizing, the per-channel power traces are transformed into a list of continuous 1s and 0s if the boundary is correctly set to be between 0 and the maximum of the amplitude. Then the pulses can be identified by classifying segments of 1s.

There are two ways to determine the gates. The first way is to use a uniform boundary, and because {\tt X} and {\tt SX} have the same duration but different amplitude, and the pulse shapes are similar to the Gaussian function and they do not have any abrupt change, their binary forms have different lengths. The type of gate can be identified by comparing the length of the segment in the binary list with the length of the binary form of the power traces of basis pulses. The second way is to use different boundaries in the search phase, i.e., firstly set a boundary between the maximum of the power traces of {\tt X} and {\tt SX}, so only {\tt X} can be found. After removing {\tt X}, then set a boundary between 0 and the maximum of the power traces of {\tt SX}. The start time can be easily computed at the same time and set to the granularity of the quantum device, where the pulses must start at multiples of the granularity.

The binarizing process is to make the method more robust under measurement noise. Another parameter for robustness is tolerance, which means the allowed length difference when comparing the length of the segment in the binary list and the length of the binary form of the power traces of the basis gate. If the difference between these two is in the range of tolerance, then it is chosen to be identified. The boundary and the tolerance are coupled in the way that the binary form of the power of one basis gate cannot be mixed with another in the range of the tolerance.

\subsubsection*{CR Attack Summary:}

Attackers are able to recover all non-virtual gates from the per-channel power traces, making this the most powerful attack among the discussed attacks. However, per-channel power trace information is needed.

\section{Defenses}
\label{sec_defenses}

In this section, we present possible defenses against quantum computer power-side channel attacks discussed in the paper.

\subsection{Preventing Timing, Total Energy, and Mean Power Attacks}
\label{sec:defense_uc}

\begin{figure}[t]
  \captionsetup[subfigure]{justification=centering}
  \centering
  \begin{subfigure}[t]{0.45\textwidth}
    \includegraphics[width=\columnwidth]{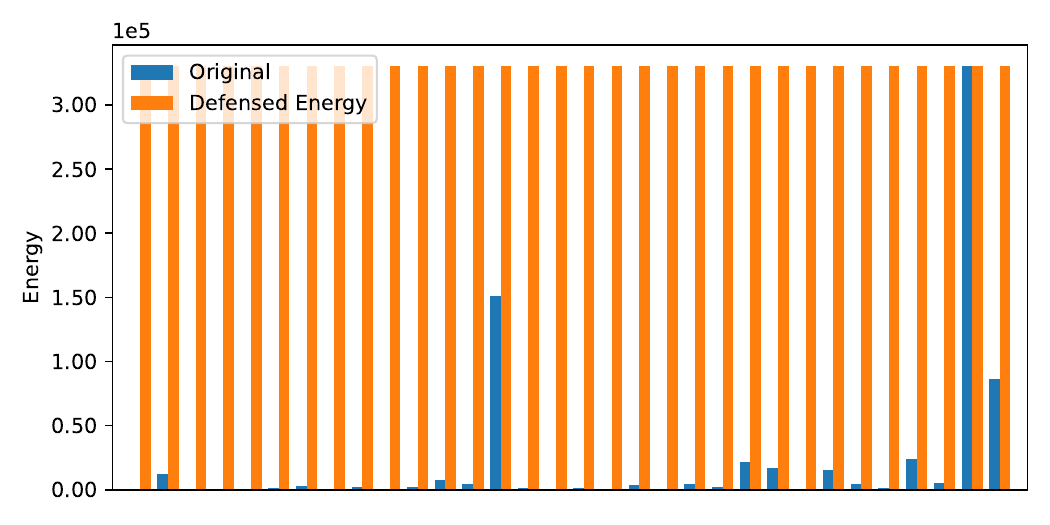}
    \caption{Energy of the benchmark.}
    \label{fig:bm_energy}
  \end{subfigure}
  
  \begin{subfigure}[t]{0.45\textwidth}
    \centering
    \includegraphics[width=\textwidth]{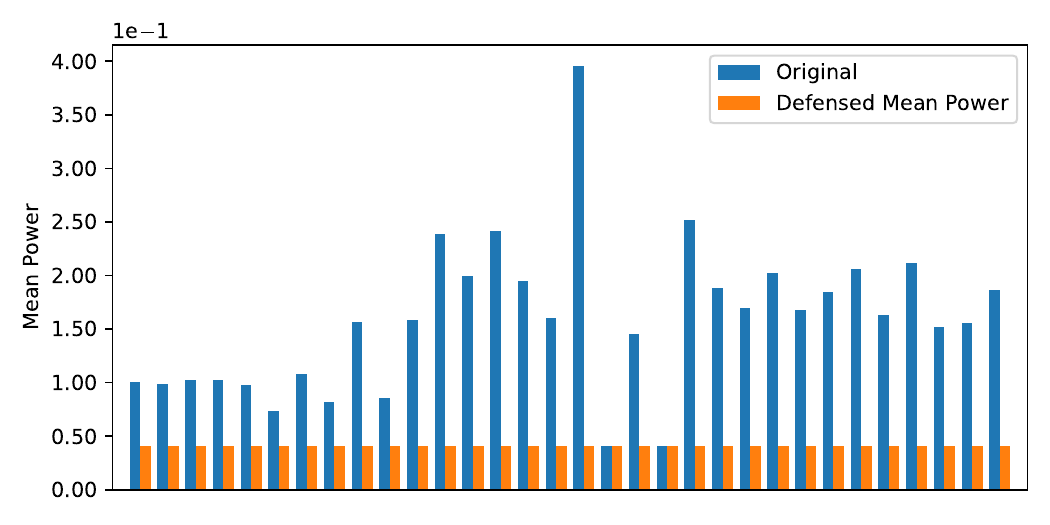}
    \caption{Mean power of the benchmark.}
    \label{fig:bm_mean_power}
  \end{subfigure}%

  \begin{subfigure}[t]{0.45\textwidth}
    \centering
    \includegraphics[width=\textwidth]{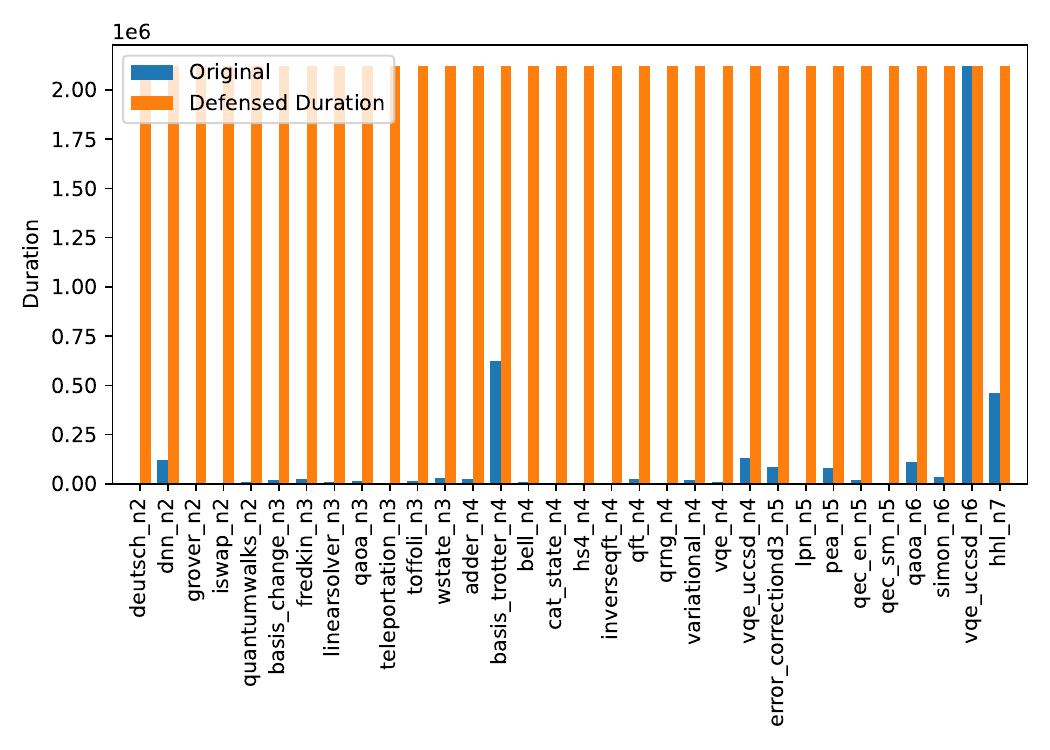}
    \caption{Duration of the benchmark.}
    \label{fig:bm_duration}
  \end{subfigure}
  \caption{\small Total energy, mean power, and timing (duration) of the circuits in the benchmarks. Blue bars show the metrics of the original circuits, and orange bars show the metrics of the circuits modified by defense methods introduced in Section~\ref{sec:defense_uc}.}
  \captionsetup[subfigure]{justification=centering}
  \label{fig:bm_info}
\end{figure}

To protect from attacks using the three scalar metrics: timing, total energy, and mean power, the insight is to add additional gates to the circuit so that the metric values of all circuits in the list can be made similar to each other. More specifically, adding gates with pulses can change the energy of the circuit, and adding gates with time can change the duration of the circuit. Because mean power is the energy divided by the duration, these two ways together with the combination of them can change the mean power of the circuits.

To defend attacks using timing, we simply choose to add delay gates. The defense is to first find the largest duration in the circuit list, and then add delay gates for all other circuits to make the duration the same as the largest duration. 

To defend attacks using energy, we choose to use two {\tt X} gates as one unit, since it is, in theory, the same as applying the identity gate and thus will not have influence in qubits. The approach is to find the largest energy in the circuit list and then add two {\tt X} gates units on different qubits to reach the largest energy. On $n$-qubit quantum devices, we only have $n$ different {\tt X} gates. The problem can be reduced to that given a list of numbers $x_1, \dots, x_n$ and a target $z$, find the combination $y_1, \dots, y_n$ that minimizes $|z - \sum_{i=1}^n x_i \cdot y_i|$. This problem can be solved with dynamic programming.

To defend attack using mean power we can also do it by adding delays. First, find the circuit with smallest mean power among the set of circuits, and then add delays to the other circuits so each can reach the smallest mean power.

The results are shown in Figure~\ref{fig:bm_info}. For the duration, because the delay gate can be with any time that is multiples of the granularity of the device, all circuits in the circuit list can be made the same duration. However, for energy, because we only have limited choices of {\tt X} gates, it is usually not able to reach the same energy for different circuits, which means the defense may not be effective with a small error rate if the circuits are well designed to avoid being protected. Similarly, circuit duration is required to be multiples of the granularity of the backends, and thus the duration usually cannot be chosen to be the duration that correctly set the mean power to be the target mean power. Nevertheless, combining with adding gates to change both the energy and duration will achieve a smaller difference from the target mean power, and this is left as future work.

\subsubsection{Defense Discussion}

In addition to the duration of the circuit which can easily be changed by adding the delay gates, gates for dynamical decoupling to mitigate qubit decoherence~\cite{viola1999dynamical} could also be added so that the duration is extended, while also better-preserving state of the qubits rather than just by using delays. Dynamical decoupling can also be utilized to change the energy of the circuit. The insertion of dynamical decoupling is already available as feature%
\footnote{\url{https://qiskit.org/documentation/stubs/qiskit.transpiler.passes.DynamicalDecoupling.html}}
in the commonly used Qiskit software development kit for working with quantum computer programs. Also, to defend from attacks using energy, four {\tt SX} gates or two {\tt CX} gates can also be added.

We note that circuits in the circuit list, such as QASMBench benchmarks, may vary a lot in terms of energy or duration, as shown in Figure~\ref{fig:bm_energy} and \ref{fig:bm_duration} shows. It is impractical to make all the metrics the same for all the circuits. To tackle this we propose two approaches. First, divide the circuits into a few groups, and make them have the same energy or duration only among circuits in a group. Second, group shots of the circuit together or cut the circuit. With the accurate reset gate, the long time for qubits to decohere to the initial states is not needed, and thus many shots can be grouped into one shot by adding reset gates after each shot. In this way, short circuits can be made to be long circuits by executing multiple shots together.%
Similarly, for very long circuits, they can be cut~\cite{PhysRevLett.125.150504, tang2021cutqc} to make short circuits so that the attacker only observes the shorter shots and does not know they belong to a longer circuit.

These above defenses are only considered for one type of side-channels. However, if different types of side-channels can be combined, then some of the defenses may be ineffective. For example, we add delays to reach the same duration, but the energy does not change. So if attackers can also measure energy, then they may still infer the circuits.

\subsection{Preventing Total Power Trace Attacks}

Protecting from attacks using power time series is more difficult since it is hard for the total power traces to be similar for all circuits without changing the functionality of circuits. However, as a feasible defense, we propose to incorporate power waster circuits into the AWG or FPGA used to generate the waveforms. Power wasters~\cite{provelengios2020power} are classical circuits that can be realized in FPGAs, the circuits use large arrays of ring oscillators to consume large amounts of power. Effectively, the total power consumed by AWG or FPGA at each time instant can be kept constant by turning power wasters on and off, so that the total power of the power wasters plus the power of the logic used to generate control pulses is constant. We note there are quantum control systems such as QICK%
\footnote{\url{https://github.com/openquantumhardware/qick}}
which already use FPGAs for control pulse generation, and a large number of research papers have studied power wasters on FPGAs, e.g.,~\cite{provelengios2020power,provelengios2020power2}.

\subsection{Preventing Per-Channel Trace Attacks}
\label{sec:defense_rz}

Per-channel traces could be defended with the power wasters, however, such defense may not be possible if FPGAs are not used for the controllers, or if there is no ability to add power waster circuits.
As a possible defense, we propose to leverage the virtual {\tt RZ} gate; this defense requires now power wasters.

{\tt RZ} gate is usually one of the basis gates in superconducting quantum computers, which rotates a single qubit around the Z axis in the Bloch sphere. While other basis gates have their calibrated pulses, {\tt RZ} gate can be implemented easily as a virtual gate with the arbitrary wave generators (AWG)~\cite{doi:10.1063/1.5089550, PhysRevA.96.022330}. If {\tt RZ} gate is implemented as a virtual gate, then it will be ``perfect'', i.e., no actual pulses are needed and thus it takes no time to execute. As we assume that the power consumption depends on the amplitudes of non-virtual pulses, {\tt RZ} gate is undetectable in power-side channels on the quantum devices where it is designed to be virtual.%

Virtual {\tt RZ} gate is valuable because any quantum gate $U$ 
can be decomposed as~\cite{PhysRevA.96.022330}:
\begin{equation}
    U(\theta, \phi, \lambda) = Z_{\phi-\pi/2} X_{\pi/2} Z_{\pi-\theta} X_{\pi/2} Z_{\lambda-\pi/2}
\end{equation}
where $Z_\theta$ is {\tt RZ} gate with the rotational angle $\theta$ and $X_{\pi/2}$ is {\tt RX} gate with rotational angle $\pi/2$, or {\tt SX} gate with a global phase. Therefore, any single-qubit gate can be realized with $X_{\pi/2}$ and {\tt RZ} gate.

To protect quantum computers from per-channel trace power-side channel attacks, we can randomly select single qubit gates $U$ in the circuit, and replace them with equivalent sequences containing the virtual {\tt RZ} gates. The modified circuit is logically equivalent to the original circuit, yet it has different non-virtual gates as well as {\tt RZ} gates for which attackers are not able to get the rotation angle from the power traces. 

We note that the {\tt RZ} gates already in the original circuit are protected from the attack, and it is the other single-gate operations we want to protect. In our implementation, we transform {\tt SX} into as number of {\tt SX} and {\tt RZ} by:
\begin{equation}
    SX = Z_{-\pi/2} \cdot SX \cdot Z_{\pi/2} \cdot SX \cdot Z_{-\pi/2}
\end{equation}
which transforms one {\tt SX} gate into two {\tt SX} gates. This protection operation can be applied recursively and thus resulting in any number of {\tt SX} gates.

The protection works as follows. If there is an {\tt SX} gate (equivalently {\tt X} gate, which can be implemented as two {\tt SX} gates) it is non-virtual and the attacker knows its rotation angle. With our defense, each {\tt SX} gate (equivalently {\tt X} gate) is replaced with an arbitrary number of {\tt SX} gates and sandwiched and inserted with {\tt RZ} gates. Therefore, if attackers retrieve a series of {\tt SX} gates from per-channel power traces, they have to guess what the original gates are composed of. The upper bound for the number of guesses can be a large number for attackers without any heuristics:
\begin{equation}
    No.\ Guesses = \sum_{i = 1}^k \sum_{j=0}^{n_i - 1} {n_i\choose j}
\end{equation}
where there are $k$ {\tt SX} sequences in the circuit, and there are $n_i$ {\tt SX} gates in the $i$-th sequence. 
This defense actually increases circuit duration very little. Applying the above transformation, except for ``qrng", which only contains one {\tt SX} gate on each qubit, the increase is less than 20\%, and less than 10\% for most of the algorithms. The increase is linear to the number of transformed {\tt SX} gates, and the number can be random and chosen considering the trade-off between security and fidelity.

Due to the limitation of native gates on the real quantum computers, we only have two real gates, {\tt SX} and {\tt X}, to participate in the above transformation. If the quantum computers provide more native gates, such as {\tt Y} gate, more transformation approaches can be implemented. More generally, it has been proved that a new circuit can be generated while only introducing a little or no experimental overhead by decomposition similar to ours~\cite{PhysRevA.94.052325}. More formally, the virtual {\tt RZ} gate decomposition scheme is to change one quantum gate $U$:
\begin{equation}
   U = U_1 \cdots U_k
\end{equation}
where at least one $U_i, i \in {1, \dots, k}$ is $RZ(\theta)$ and $U$ and $U_{i_1} \cdots U_{i_n}$ are not equivalent. By modifying the circuit and replacing randomly selected gates with equivalent gate sequences that contain {\tt RZ} gates, attackers are not able to reconstruct the original circuit fully from the power traces since they do not know where the virtual {\tt RZ} gates are, and what are the rotation angles.

\subsection{Defenses using Custom Gates}

If the custom gates, for which users can specify their own pulses, are supported, then there would be additional defensive possible. To protect from attacks using energy, mean power, or duration, custom pulses can be added to change the energy and mean power, and thus it is possible to make all circuits in the list have the same energy or mean power. For power time series, the custom pulses can behave like power masks and thus it is also possible for all the circuits to have the same power time series. In addition, for circuit reconstruction, though the attack may be able to differentiate custom pulses from native pulses from the power traces, the attacker cannot know the functionality of the custom pulses, and thus the circuits can be protected.

\section{Discussion and Future Work}

Calibrations are usually done automatically, and due to physical conditions of qubits and experiment errors, calibration data is changed over time. Therefore, if attackers do not have the calibration data of when the quantum computer was calibrated, they may introduce errors in the inference and thus it has a higher probability to make a wrong guess. Figure~\ref{fig:accuracy} showed how accuracy of the attacks degrades with higher error rates.

Considering the virtual {\tt RZ} gate, there is still some computation necessary in the AWG to shift the phase. This computation may cause some power consumption or timing difference in the AWG or FPGA used to generate the control pulses. For the current work, we assume this small computation is not noticeable in the power traces, compared to the real pulse generation logic. However, if whether {\tt RZ} gates are applied and what are the angles for them can be leaked in some way, attackers can be more powerful in our discussed situations.

Lastly, because different quantum circuits will have different features, these can be utilized together with side-channels. For exmaple, {\tt CX} gates' relative locations and their operating qubits may be a useful feature to identify circuits. If attackers can pinpoint the locations and operating qubits of {\tt CX} gates in a circuit, then they may be able to identify the circuit. Developing heuristics to help attackers is left as future work.

\section{Conclusion}
\label{sec_conclusion}
This work presented the first exploration of side-channel attacks on quantum computers. We propose the threat model and several applications of quantum computer side-channels, and evaluate how effective power traces can be used in the various cases. As this work shows, side-channels attacks could be powerful and practical for inferring secret information about circuits executing on quantum computers, and appropriate defenses need to be deployed.

\balance

\bibliographystyle{plain}
\bibliography{bibliography}

\end{document}